\documentclass[lettersize,journal]{IEEEtran}
\usepackage[utf8]{inputenc}
\usepackage[cmex10]{amsmath}
\usepackage{upgreek}
\usepackage{booktabs}
\usepackage{epsfig}
\usepackage{latexsym}
\usepackage{multirow}
\usepackage{stfloats}
\usepackage{epstopdf}
\usepackage{color}  
\usepackage{tabularx} 
\usepackage{algorithm}
\usepackage{algpseudocode}
\usepackage{amssymb}
\usepackage{enumerate}
\usepackage{array}
\graphicspath{{./Figures/}}
\usepackage{color}
\usepackage{bbm}
\usepackage{bm}
\usepackage{cite}
\usepackage[tight,footnotesize]{subfigure}
\usepackage{balance}
\usepackage{mathrsfs}
\usepackage{verbatim}
\usepackage{dsfont}
\usepackage{verbatim}
\usepackage{setspace}
\usepackage{diagbox}
\usepackage{multicol}
\usepackage{environ}
\usepackage{tikz}
\usepackage{amsmath}
\usepackage{stfloats}
\usepackage{algorithm}
\usepackage{algpseudocode}
\usepackage{amsmath}
\usepackage{graphics}
\usepackage{epsfig}
\usepackage{caption}
\allowdisplaybreaks[4]
\usepackage{amsmath}
\usepackage{amsthm}
\usepackage{authblk}
\newcommand{\wjg}[1]{\textcolor{black}{{#1}}}

\newcounter{TempEqCnt}
\def\d{\ensuremath{\mathrm{d}}}
\def\BibTeX{{\rm B\kern-.05em{\sc i\kern-.025em b}\kern-.08em
		T\kern-.1667em\lower.7ex\hbox{E}\kern-.125emX}}
\begin{document}
\title{Attenuation and Loss of Spatial Coherence Modeling for Atmospheric Turbulence in Terahertz UAV MIMO Channels
}
%\title{Atmospheric Turbulence in Terahertz UAV MIMO Channels: Attenuation and Loss of Spatial Coherence Modeling}
\author{Weijun Gao,~\IEEEmembership{Graduate Student Member,~IEEE}, Chong Han,~\IEEEmembership{Senior~Member,~IEEE}, Zhi Chen,~\IEEEmembership{Senior Member,~IEEE}
\thanks{
Weijun~Gao is with the Terahertz Wireless Communications (TWC) Laboratory, Shanghai Jiao Tong University, Shanghai 200240 China (email:~gaoweijun@sjtu.edu.cn). 

Chong~Han is with the Terahertz Wireless Communications (TWC) Laboratory, Department of Electronic Engineering and Cooperative Medianet Innovation Center (CMIC), Shanghai Jiao Tong University, Shanghai 200240 China (email:~chong.han@sjtu.edu.cn). 

Zhi~Chen is with with the National Key Laboratory of Science and Technology on Communications (NCL), University of Electronic Science and Technology of China, Chengdu 611731, China (email: chenzhi@uestc.edu.cn).}
}

%\markboth{IEEE Transactions on Wireless Communications, revised in January 2024}{}
	\maketitle
	\thispagestyle{empty}
	\begin{abstract}
        Terahertz (THz) wireless communications have the potential to realize ultra-high-speed and secure data transfer with miniaturized devices for unmanned aerial vehicle (UAV) communications. 
        %Existing THz UAV channel models assume a homogeneous medium. 
        The atmospheric turbulence due to random airflow leads to spatial inhomogeneity of the communication medium, which is yet missing in most existing studies, leading to additional propagation loss and even loss of spatial coherence (LoSC) in MIMO systems. 
        %Even worse, a unique influence of turbulence, namely, the loss of spatial coherence for the THz ultra-massive MIMO systems, arises that is not analyzed in most existing studies. 
        In this paper, the attenuation and loss of spatial coherence for atmospheric turbulence are modeled in THz UAV MIMO channels. 
        Specifically, the frequency- and altitude-dependency of the refractive index structure constant (RISC), as a critical statistical parameter characterizing the intensity of turbulence, is first investigated. 
        Then, the LoSC, fading, and attenuation caused by atmospheric turbulence are modeled, where the turbulence-induced fading is modeled by a Gamma-Gamma distribution, and the turbulence attenuation as a function of altitude and frequency is derived. 
        %Numerical simulations on the RISC, fading, and attenuation are presented to quantitatively analyze the influence of turbulence for the THz UAV channels. 
        Numerical results show that the turbulence leads to at most 10\,dB attenuation with frequency less than 1\,THz and distance less than 10\,km. Furthermore, when the distance is 10\,km and the RISC is $10^{-9}\,\textrm{m}^{-2/3}$, the loss of spatial coherence effect leads to 10\,dB additional loss for a $1024\times1024$ ultra-massive MIMO system.
	\end{abstract}
	
	\section{Introduction}
	With the increasing demand for immersive and hyper-reliable wireless communications, Terahertz (0.1-10\,THz) communications have attracted great research attention for 6G and beyond wireless systems~\cite{han2022terahertz}. Thanks to the sub-millimeter wavelength and multi-tens-of-GHz continuous bandwidth, the THz band can support ultra-fast, highly directional, and secure wireless links.
    One of the potential applications is the wireless communications between unmanned aerial vehicles (UAV), such as UAV-aided ubiquitous coverage and UAV-based relaying\cite{khuwaja2018survey,li2021ray}.
    First, UAV communication services bolster transmissions of high-resolution images or low-latency commands. Second, miniaturized communication devices enabled by the short wavelength of THz band can lighten the load of small UAVs. Third, the high directivity and the enhanced communication security can effectively prevent the UAVs from being eavesdropped through the UAV channels with bare obstructions~\cite{bithas2020uav}.
	
    Modeling of the THz UAV channel is one of the most fundamental topics in studying THz wireless communications. 
    \wjg{Existing studies on THz UAV channel modeling focus on air-to-ground channels\cite{cui2022cluster,cui2019measurement,an2022measurement}, where the multi-path effects reflected by various terrains are investigated. For the UAV-to-UAV scenarios,} due to the lack of multi-paths in aerial communication scenarios, the THz UAV channel is reasonably considered to transmit through the line-of-sight (LoS) path between the transceivers.
    In prior studies~\cite{li2021ray,jornet2011channel}, the LoS signal attenuation consists of free-space path loss, the molecular absorption mainly due to water vapor, and the scattering effect of raindrops and dust. 
    These channel models assume that the propagation medium is static, while the inhomogeneity of propagation medium led by chaotic airflow, i.e., atmospheric turbulence, is not considered. 
    
    In last several decades, the turbulence-induced fading and attenuation of wireless communications has been analyzed for optical or radar communications, e.g., laser beam propagation~\cite{andrews2005laser,esmail2021experimental}. 
    Since the turbulence is deterministically governed by the Navier-Stokes equations~\cite{temam2001navier}, the modeling of turbulence-induced fading and attenuation meets mathematical difficulty in solving this non-linear equation. As a result, only statistical models are accessible to characterize the turbulence.
    Many statistical fading models of irradiance of laser communications in turbulent media are analyzed, among which the Gamma-Gamma distribution is used to model the irradiance fluctuation in~\cite{wilfert2004statistical,al2001mathematical,arya2018non}. 
    Attenuation models for laser communications in turbulent media are developed and summarized in~\cite{dordova2010calculation}.     
    Despite the research effort on atmospheric turbulence in optical communications, it is rarely studied for micro-wave or millimeter-wave communications for the following two reasons. 
    First, it is discovered that the Rytov variance, as a critical parameter to characterize the influence of turbulence, is proportional to $f^{7/6}$, where $f$ represents the frequency~\cite{dordova2010calculation}. Therefore, the influence of turbulence is small in these low-frequency bands. 
    Second, the turbulence only leads to severe attenuation for high directional communications\cite{andrews2005laser}, and omnidirectional communications in a rich scattering environment is not significantly affected by turbulence.
    Since the THz band is located between the millimeter-wave band and the optical band and usually high directional, it is essential to investigate its influence on THz UAV communications.

    Previous studies on the influence of turbulence on THz communications extend the conclusions and results in optical band to the THz band. 
    Since the directional THz beam propagation is analogous to laser propagation, their conclusions and takeaway lessons is reasonable.
    The additional propagation loss and fading effects caused by turbulence are analyzed in~\cite{ma2016terahertz}. In~\cite{ma2015experimental}, the effects of turbulence on THz links are experimentally studied under controlled lab conditions.
    In~\cite{cang2019impact}, the amplitude and phase fluctuation of THz signals due to turbulence is analyzed.
    
    One of the limitations of these works is the assumption that the transceivers are equipped with a single high gain directional antenna. 
    \wjg{To overcome the distance limitation, the ultra-massive multiple-input multiple-output (UM-MIMO) communication system is widely assumed in the literature\cite{akyildiz2016realizing,yan2020dynamic,yi2023full,chen2023hybrid}.}
    Since the analysis for laser beam propagation assumes single antenna, the extended analysis in the THz band is not valid in THz UM-MIMO systems.
    Moreover, the frequency- and altitude-dependent statistical parameter for turbulence, i.e., refractive index structure constant (RISC), is based on the measurement data in optical band. Therefore, the RISC in the THz band needs to be re-investigated.

    Due to the aforementioned challenges, 
    we analyze the influence of atmospheric turbulence on THz UAV MIMO channels.
    A unique phenomenon led by turbulence known as loss of spatial coherence (LoSC) is first observed and analyzed in UM-MIMO. The UM-MIMO channel is composed of lots of single-input single-output (SISO) channel between each Tx and Rx antenna.  
    In non-turbulent media, two different SISO channels for near Tx and Rx antennas are perfectly correlated, which means the SISO channels in the UM-MIMO can be constructively combined by well designing the amplitudes and phases. 
    This correlation due to the near locations of Tx and Rx is known as the spatial coherence of signal transmission in UM-MIMO.
    Turbulent media is demonstrated to be able to destroy the spatial coherence by introducing uncorrelated random amplitude and phase fluctuation in different SISO channels in UM-MIMO.
    We analyze the model this LoSC effect in THz UAV MIMO channels.  
    Moreover, the attenuation model based on the extended RISC model is re-investigated. 
    The contributions of this paper are summarized as follows.

    %equipped with high gain reflector antennas~\cite{taherkhani2020performance}. MIMO is not considered.

    %For the aforementioned motivations, it is necessary to analyze and model the effect of atmospheric turbulence in THz UAV communications.    

    %In this paper, we statistically model the fading and attenuation effect caused by atmospheric turbulence for THz UAV channels. 
    
    %Specifically, we first analyze the RISC (RISC), which is a key parameter to characterize turbulence. We develop the model of RISC at different frequencies and altitudes in the THz band based on the statistical turbulence model in the infrared frequency band. Second, we model the fading caused by turbulence by using a Gamma-Gamma distribution, which is a universal model applicable to the turbulence of various intensities. Finally, the fading and attenuation caused by turbulence at different propagation distances and RISC are evaluated to quantitatively demonstrate the influence of turbulence on THz UAV channels. 

    %The detailed contributions of this paper are summarized as follows.
    \begin{itemize}
        \item\textbf{We develop a system model for the THz UM-MIMO communications in turbulent media.} According to the Kolmogorov turbulence theory, we propose a RISC model at different frequencies and altitudes in the THz band extended from the statistical turbulence model in the optical band.
        \item\textbf{The LoSC of the UM-MIMO system led by atmospheric turbulence is analyzed and modeled.} 
        By developing a closed-form expression on the ergodic capacity in turbulent media of the THz MIMO system, the effect of LoSC is demonstrated and characterized by normalized covariance between different SISO channels. 
        Then,  we derive a closed-form expression for the attenuation due to LoSC based on the statistical model of turbulence-induced refractive index fluctuation.
        \item\textbf{The fading and attenuation model led by turbulent media in the THz band is established.} 
        %Four statistical fading models are analyzed, based on which the attenuation model for the THz UAV channels under various conditions is described. 
        We model the fading caused by turbulence by using a Gamma-Gamma distribution, which is a universal model applicable to the turbulence of various intensities. The attenuation model is then established and studied based on fading model by the Andrew's method.
        \item \textbf{We explore how different parameters can influence the attenuation and LoSC in the THz UAV UM-MIMO system.} 
        The LoSC, fading, and attenuation of atmospheric turbulence at different distances, RISC, and array size are evaluated to quantitatively demonstrate the influence of turbulence on THz UAV channels.
        Extensive numerical results demonstrate that when the distance is $10\,\textrm{km}$ and the RISC is $10^{-9}\,\textrm{m}^{-2/3}$, the LoSC effect leads to approximately $10\,\textrm{dB}$ additional loss for \wjg{an UM-MIMO with $32\times 32$ uniform planar arrays at the transmitter and receiver side\cite{akyildiz2016realizing,chen2023hybrid,yan2020dynamic}}, and the turbulence can lead to at most $10\,\textrm{dB}$ attenuation with distance less than $10\,\textrm{km}$. 
    \end{itemize}

The remainder of the paper is organized as follows. In Sec.~\ref{sec:sys}, the THz UM-MIMO system model in turbulent media is described and modeled, where the atmospheric turbulence in the THz band is statistically characterized according to Kolmogorov's theory. In Sec.~\ref{sec:LoSC}, the LoSC effect caused by turbulent media is illustrated and analyzed, and the ergodic capacity of UM-MIMO system in turbulent media and the additional loss due to LoSC are formulated and derived.
In Sec.~\ref{sec:fading_and_attenuation}, the fading and attenuation effects of turbulence on the THz UAV communications are investigated. Numerical evaluations for the altitude-dependent RISC, Rytov variance, LoSC, fading, and attenuation are evaluated in Sec.~\ref{sec:numerical} to quantitatively measure the influence of atmospheric turbulence on THz communications. The paper is concluded in Sec.~\ref{sec:concl}.

\textit{Notation:} 
$x$ is a scalar. $\mathbf{x}$ represents a vector. $\mathbf{X}$ stands for a matrix. $x_{ij}$ denotes the element at the $i^{\textrm{th}}$ row and the $j^{\textrm{th}}$ column in $\mathbf{X}$. $x^*$ denotes the conjugate. $\mathbf{X}^{\textrm{H}}$ represents the Hermitian transpose. $\exp(\cdot)$ defines the exponential function. $\textrm{Re}(\cdot)$ returns the real part of a complex number. $|x|$ denotes the L2-norm.
\section{Terahertz UAV UM-MIMO System Model in Turbulent Media}~\label{sec:sys}
\begin{figure*}
    \centering
    \includegraphics[width=0.8\textwidth]{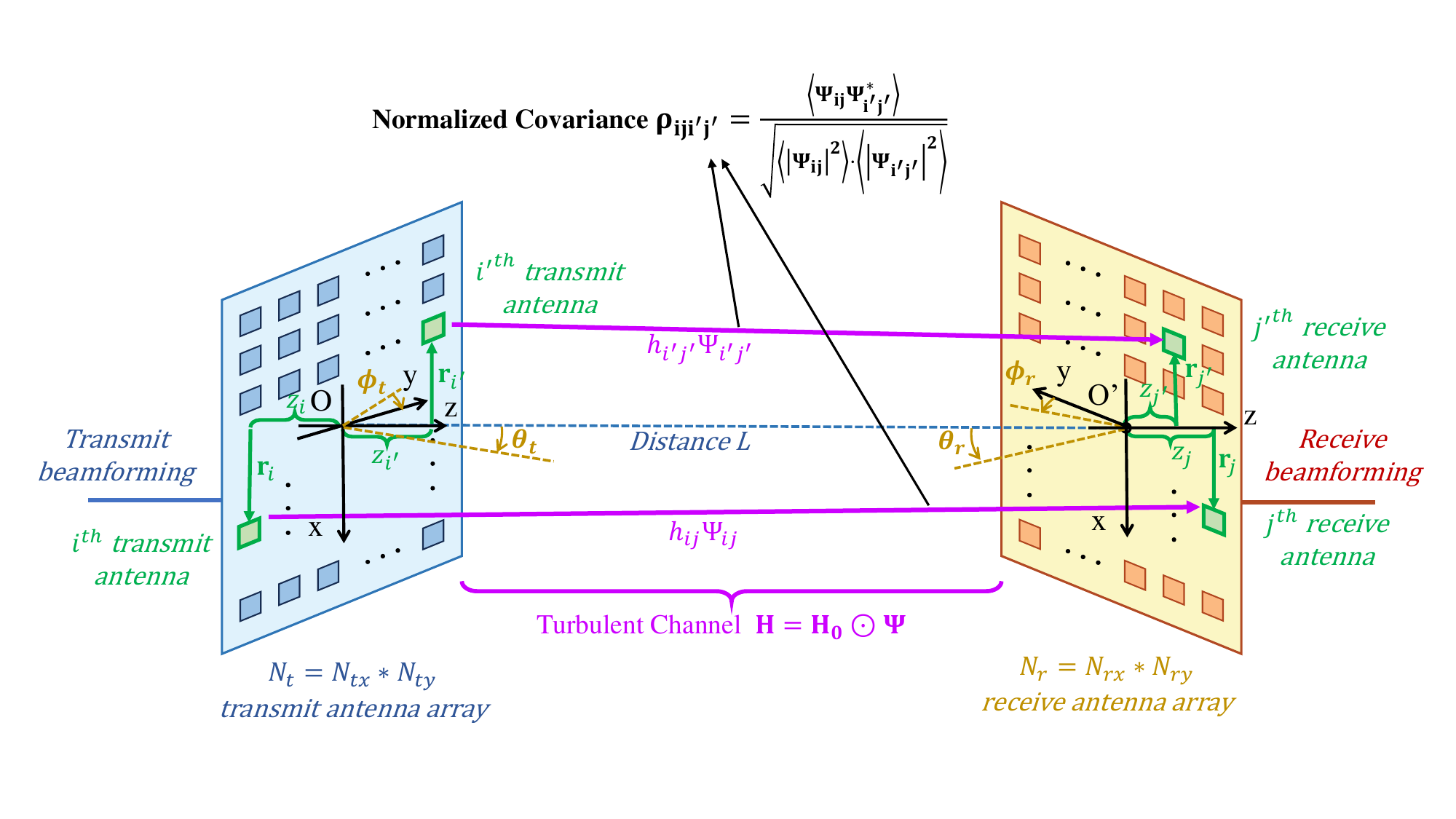}
    \captionsetup{font={footnotesize}}
    \caption{System Model: MIMO in turbulent media.}
    \label{fig:system_model}
\end{figure*}  
We consider a THz UM-MIMO system in UAV turbulent channel as depicted in Fig.~\ref{fig:system_model}, where a transmitter equipped with $N_t=N_{tx}\times N_{ty}$ antennas transmits the signal to the receiver (Rx) equipped with $N_r=N_{rx}\times N_{ry}$ antennas.
At the transmitter (Tx) side, the signal $\mathbf{s}$ is first precoded by the beamforming matrix $\mathbf{p}=[p_1,p_2,\cdots,p_{N_t}]^T\in\mathcal{C}^{N_t\times 1}$ as
	\begin{equation}
		\mathbf{x}=\mathbf{p}s,
		\label{eq:x}
	\end{equation}
where \wjg{$s\in\mathcal{C}$ represents one of the complex symbols in the symbol stream $\mathbf{s}$ transmitted to the receiver side}, $\mathbf{x}=[x_1,x_2,\cdots,x_{N_t}]^T\in\mathcal{C}^{N_t\times 1}$ stands for the transmitted signal.
We assume that the baseband signal has normalized power, i.e., $E[\mathbf{s}\mathbf{s}^{\textrm{H}}]=P_{t}$, where $(\cdot)^{H}$ denotes the Hermitian transpose and $P_{t}$ is the transmit power.
We construct a Cartesian coordinate where the origin is at the center of the transmit antenna array, the z-axis is along the LoS path as depicted in Fig.~\ref{fig:system_model}. The direction of departure (DoD) and direction of arrival (DoA) are denoted by the angle pairs $(\theta_{t},\theta_{r})$ and  $(\phi_{t},\phi_{r})$, where $\theta$ and $\phi$ represents the elevation and azimuth angle, respectively. The location of the $i^{\textrm{th}}$ antenna at Tx and the $j^{\textrm{th}}$ antenna at Rx can be expressed by $(\mathbf{r}_i,z_i)$ and $(\mathbf{r}_j,z_j)$, respectively, where the vector $\mathbf{r}_i$ and $\mathbf{r}_j$ denote their locations on the x-y plane and $z_i$ and $z_j$ stand for the distance along the z-axis.
By assuming a combining matrix $\mathbf{w}=[w_1,w_2,\cdots,w_{N_r}]^T\in\mathcal{C}^{N_r\times 1}$ at the receiver side, the received signal is expressed as
	\begin{equation}
		y=\mathbf{w}^{\textrm{H}}\mathbf{H}\mathbf{x}+\mathbf{w}^{\textrm{H}}\mathbf{n},
	\end{equation}
	where $\mathbf{H}\in\mathcal{C}^{N_r\times N_t}$ denotes the THz turbulent channel, and $\mathbf{n}\in\mathcal{C}^{N_r\times 1}$ represents the additive white Gaussian noise (AWGN).
    We assume that the THz UAV channel includes only the LoS path due to the high spreading loss and reflection loss of THz wave propagation. Therefore, by denoting $\mathbf{\Psi}\in\mathcal{C}^{N_r\times N_t}$ represents the turbulence-induced channel perturbation matrix, the THz channel for turbulent media is given by
	\begin{equation}
		\mathbf{H}=\mathbf{H}_0\odot \mathbf{\Psi},
  \label{eq:psi}
	\end{equation}
	where $\mathbf{H}_0\in\mathcal{C}^{N_r\times N_t}$ denotes the THz MIMO channel in non-turbulent media \wjg{given by~\cite{yan2020dynamic}}
	\begin{equation}
		\mathbf{H}_0=\alpha_{\textrm{LoS}}\mathbf{a}_r(\theta_{r},\phi_{r})\mathbf{a}_t^{\textrm{H}}(\theta_{t},\phi_{t}),
		\label{eq:H0}
	\end{equation}
where $\mathbf{a}_t(\phi_{t},\theta_{t})\in \mathcal{C}^{N_t\times 1}$ and $\mathbf{a}_r(\phi_{r},\theta_{r})\in \mathcal{C}^{N_r\times 1}$ represent the array steering vectors at Tx and Rx, respectively. 
$\alpha_{\textrm{LoS}}$ denotes the path gain of the LoS path in non-turbulent media, and $\Psi_{ij}$ stands for turbulence perturbation on the channel from the $i^{\textrm{th}}$ transmit antenna to the $j^{\textrm{th}}$ receive antenna.
\subsection{Absorption and Scattering Model in Terahertz UAV Channel}
Besides free-space spreading loss, THz LoS signals experience absorption and scattering loss due to the interaction between the \textcolor{red}{electromagnetic (EM)} wave and medium particles. 
On one hand, molecules such as water vapor and oxygen can be excited by the THz wave and absorb part of the energy of the EM wave, which is referred to as the molecular absorption effect~\cite{jornet2011channel}. 
On the other hand, small particles caused by extreme weather like rain, snow, dust, and fog can scatter the EM wave and lead to scattering loss. 
The THz UAV channel impulse response caused by free-space path loss, absorption, and scattering in non-turbulent media can be expressed by~\cite{jornet2011channel, ITUR}
\begin{align}
    \alpha_{\textrm{LoS}} =\frac{c}{4\pi fL} \exp\left[-\frac{1}{2}k_{\textrm{abs}}(f,h)L-\frac{1}{2}k_{\textrm{sca}}(f,h)L\right],
    \label{eq:alos}
\end{align}
where $c$ stands for the speed of light, $L$ denotes the propagation distance, $\frac{c}{4\pi fL}$ represents the free-space path loss, and $k_{\textrm{abs}}(f,h)$ and $k_{\textrm{sca}}(f,h)$ stand for the absorption and scattering  coefficient, respectively. The molecular absorption coefficient $k_{\textrm{abs}}$ is characterized in~\cite{jornet2011channel}, which is a frequency- and altitude-dependent. $f$ stands for the frequency and $h$ is the altitude. Water vapor dominates the molecular absorption with six orders of magnitude higher than oxygen and others, and thus we can express $k_{\textrm{abs}}(f,h)$ in~\eqref{eq:alos} as
	\begin{equation}
	\begin{aligned}
	    k_{\textrm{abs}}(f,h)\approx k_{\textrm{water},\textrm{terr}}(f)\alpha_{\textrm{water}}(h),
	\end{aligned}
	\end{equation}
where $k_{\textrm{water},\textrm{terr}}(f)$ denotes the terrestrial water vapor absorption coefficient and $\alpha_{\textrm{water}}(h)$ represents the ratio of water vapor density at altitude $h$ to the terrestrial one.
	
The scattering effect between the EM wave and a certain type of particle can be classified into two cases, \wjg{namely Rayleigh scattering and Mie scattering~\cite{yang2023universal}}, depending on the relationship between the wavelength and the size of the particle. If the wavelength is larger than the size of the particle, the scattering is Rayleigh scattering, or otherwise, it becomes Mie scattering. At THz frequencies, the wavelength at millimeters or sub-millimeters is smaller than the \textcolor{red}{radius} of common particles such as rain, fog, and snow on the order of $10-100\,\textrm{mm}$~\cite{ma2018effect}, and therefore the scattering effect in the THz band is mainly Mie scattering. The Mie scattering loss coefficient $k_{\textrm{sca}}$ in dB/km in~\eqref{eq:alos} can be represented by~\cite{wilfert2004statistical}
	\begin{equation}
	    k_{\textrm{sca}}(f,h)=4.343\int_{0}^{\infty}\sigma_{\textrm{ext}}(f,r)N(h,r)dr,
	\end{equation}
where $r$ denotes the radius of the particle, and $N(h,r)$ represents the molecular size distribution at altitude $h$ commonly modeled by an exponential distribution $N(h,r)=N_0(h)\exp[-\rho_0(h) r]$. $N_0(h)$ and \wjg{$\rho_0(h)$} are altitude-dependent coefficients. $\sigma_{\textrm{ext}}(f,r)$ stands for the extinction cross-section in Mie theory~\cite{norouzian2019rain}, which can be expressed by
\begin{equation}
	\begin{aligned}
        \sigma_{\textrm{ext}}(f,r)
        =&\frac{2\pi}{k^2}\sum_{m=1}^{\infty}(2m+1)\textrm{Re}[a_m(f)+b_m(f)]\\
        \approx&\frac{2\pi}{k^2}\sum_{m=1}^{x+4x^{1/3}+2}(2m+1)\textrm{Re}[a_m(f)+b_m(f)],
	\end{aligned}
\end{equation}
where $k=\frac{2\pi f}{c}$ is the wave number, and the threshold parameter is given by $x=\frac{2\pi r}{\lambda}$, where $\lambda$ represents the wavelength. 
$a_m(f)$ and $b_m(f)$ represent Mie scattering coefficients. \wjg{Since it is impractical to sum from the zero-order term to infinite-order term, we apply the approximation with the maximum order $x+4x^{1/3}+2$~\cite{norouzian2019rain}}.
	\begin{figure}
		\centering
\includegraphics[width=0.48\textwidth]{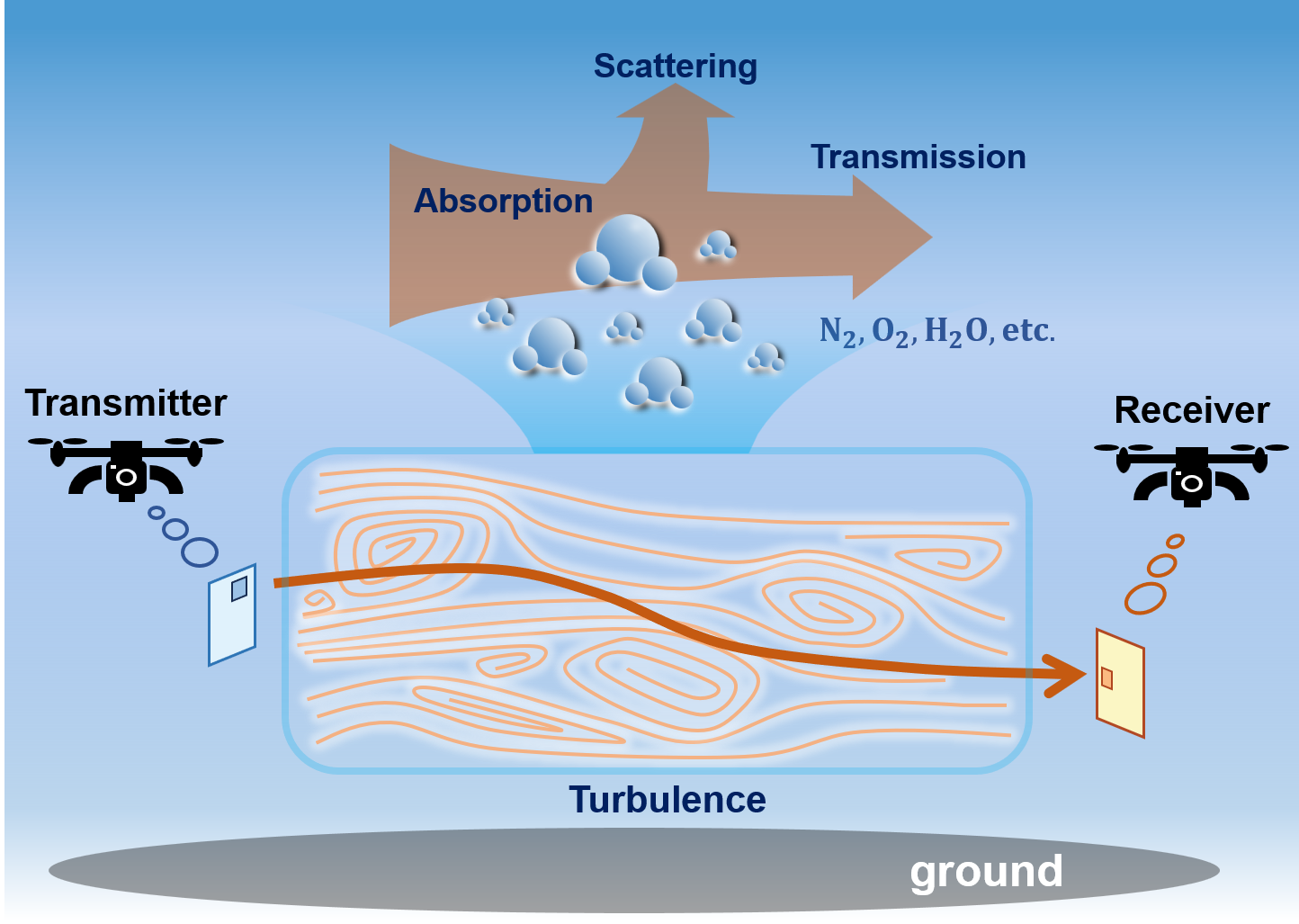}
		\captionsetup{font={footnotesize}}
    \caption{Terahertz UAV channel model in turbulent media.}
	    \label{fig:system_model_uav}
	\end{figure}  
\subsection{Statistical Model of Turbulent Media}
The refractive index of the turbulent media is not uniform and can randomly distort the transmitted signal, as shown in Fig.~\ref{fig:system_model_uav}.
Precisely characterizing the turbulent flow is critical in analyzing its influence on THz wave propagation.
Navier-Stokes equations, as fundamental to describe the motion of viscous fluid in hydrodynamics, are the most straightforward way to model atmospheric turbulence. However, due to the difficulty in mathematically solving those nonlinear equations, it is impractical to deterministically solve them to model the effect of turbulence on wave propagation. 

Instead, Kolmogorov's theory is widely used to model the turbulence as the most fundamental statistical turbulence theory~\cite{kraichnan1964kolmogorov}. It is assumed that turbulent media can be regarded as a lot of small unstable air masses called eddies and the feature of each eddy is statistically isotropic and homogeneous. The size of such eddies ranges from an inner turbulent scale $l_0$ to an outer turbulent scale $L_0$. According to dimension analysis and the law of conservation of energy, it is discovered that the temperature and refractive index of two points $i$ and $j$ satisfy the 2/3 power law~\cite{kraichnan1964kolmogorov} that 
     \begin{equation}
         \langle (T_i-T_j)^2\rangle=
         \begin{cases}
              C_T^2 L_{ij}^{2/3}, &l_0<L_{ij}<L_0,\\
              C_T^2 l_0^{-4/3}L_{ij}^2, &L_{ij}\le l_0,
         \end{cases}
     \label{eq:DTT}
     \end{equation}
     \begin{equation}
        \langle (n_i-n_j)^2\rangle=
         \begin{cases}
              C_n^2 L_{ij}^{2/3}, &l_0<L_{ij}<L_0,\\
              C_n^2 l_0^{-4/3}L_{ij}^2, &L_{ij}\le l_0,
         \end{cases}
         \label{eq:Dnn}
     \end{equation}
     where $\langle\cdot\rangle$ represents the statistical mean \wjg{over time}. $T_i$ and $T_j$ represent the temperatures of points $i$ and $j$.  $n_i$ and $n_j$ represent the refractive indices of points $i$ and $j$, respectively. $L_{ij}$ denotes the distance between $i$ and $j$. $C_T^2$ and $C_n^2$ stand for the coefficients in~\eqref{eq:DTT} and~\eqref{eq:Dnn} in $\textrm{K}^2\textrm{m}^{-2/3}$ and $\textrm{m}^{-2/3}$, respectively, which are known as the temperature structure index and the RISC.

A three-dimensional spatial power spectrum of the refractive index parameter is widely used to characterize the turbulence-induced refractive index fluctuation. By assuming two locations denoted by the position vectors $\mathbf{r}_1$ and $\mathbf{r}_2$ \wjg{with the origin $O'$ in Fig.~\ref{fig:system_model}}, the spatial power spectrum is defined by
\begin{equation}
	\Phi_n(\mathbf{k})\triangleq\frac{1}{(2\pi)^3}\iiint_{-\infty}^{\infty}\exp(-i\mathbf{k}\cdot\Delta\mathbf{r})\langle n(\mathbf{r}_1)n^{*}(\mathbf{r}_2)\rangle\d^3 \mathbf{r},
\end{equation}
where $\Delta\mathbf{r}=\mathbf{r}_1-\mathbf{r}_2$, $\mathbf{k}$ denotes the wave number vector.
Based on the Kolmogorov's turbulence theory, the spectrum is given by~\cite{andrews2005laser}
\begin{equation}
\Phi_n(\mathbf{k})=0.033C_n^2\kappa^{-11/3},\quad 1/L_0\ll \kappa\ll 1/l_0
	\label{eq:Phi_Kolmo}
\end{equation}
where $\kappa=|\mathbf{k}|$ is the scalar number of $\mathbf{k}$.
According to~\eqref{eq:Phi_Kolmo}, $C_n^2$ is a critical statistical parameter to characterize the turbulence.
\subsection{Generalized Hufnagle-Valley Model for RISC in the Terahertz Band}
In the infrared and higher frequency bands, the RISC at different altitudes $C_{n,\textrm{infrared}}^2(h)$ with wavelength around $0.5\,~\mu m$ can be modeled according to the Hufnagle-Valley model~\cite{tyson1996adaptive} as
     \begin{equation}
     \begin{aligned}
             C_{n,\textrm{infrared}}^2(h)&=0.00594(v/27)^2(10^{-5}h)^{10}e^{-\frac{h}{1000}}\\
             &+2.7\times 10^{-16}e^{-\frac{h}{1500}}+Ae^{-\frac{h}{100}},
     \end{aligned}
     \label{eq:RISC}
     \end{equation}
     where $v$ represents the average wind velocity in $\textrm{m}/\textrm{s}$, and $A$ is the terrestrial RISC.
     However, the RISC model in the THz band has not been investigated due to the lack of measurement data. Instead, we propose a generalized Hufnagle-Valley model for the RISC in the THz band based on the relationship between refractive index and temperature. Since temperature does not depend on frequency, its structure constant is frequency-invariant according to the definition~\eqref{eq:DTT}. Therefore, the temperature structure constant in the THz band is equal to that in the infrared band, i.e., $C_{T,\textrm{THz}}^2=C_{T,\textrm{infrared}}^2$.
     Then, given the relationship between the refractive index and temperature in the two frequency bands $n_{\textrm{THz}}(T)$ and $n_{\textrm{infrared}}(T)$ as~\cite{federici2010review}
     \begin{align}
         n_{\textrm{THz}}(T)&=1+77.6\times10^{-6}\left(\frac{P_a}{T}+4810\frac{P_v}{T^2}\right),
         \label{eq:nTHz}
         \\
         n_{\textrm{infrared}}(T)&=1+79\times 10^{-6}\frac{P_a}{T},
         \label{eq:n_infra}
     \end{align}
     where $P_a$ and $P_v$ denote the pressure of atmosphere and water vapor in millibar, respectively, we can express the RISC model in the THz band as
\begin{equation}
\begin{aligned}
C_{n,\textrm{THz}}^2(h)&=C_{n,\textrm{infrared}}^2(h)*\Big(\frac{\partial n_{\textrm{THz}}(T)}{\partial T}\Big)^2\Big/\Big(\frac{\partial n_{\textrm{infrared}}(T)}{\partial T}\Big)^2\\
&\approx C_{n,\textrm{infrared}}^2(h).
\end{aligned}
         \label{eq:cn2_transform}
     \end{equation}
     By substituting~\eqref{eq:nTHz} and~\eqref{eq:n_infra} into~\eqref{eq:cn2_transform}, we can transform the RISC model in the infrared frequency band into a RISC model in the THz band according to~\eqref{eq:cn2_transform}. 

\section{Loss of Spatial Coherence in THz UM-MIMO system}~\label{sec:LoSC}
Unlike directional antenna, UM-MIMO relies on the spatial coherence among antennas to create directivity. 
For example, the phases of a transmit antenna array are designed such that their signals constructively interfere at the receiver, i.e., $\mathbf{p}^{\textrm{opt}}=\mathbf{a}_t(\theta_{t},\phi_{t})$. Similarly, the receiver constructively combines the signals received from different antennas by adopting maximum ratio combining, i.e.,  $\mathbf{w}^{\textrm{opt}}=\mathbf{a}_r(\theta_{r},\phi_{r})$.
\wjg{However, due to the random amplitude and phase perturbation introduced by the atmospheric turbulence, the channel capacity of turbulent channel is computed by the ergodic capacity $\bar{C}$. Based on~\eqref{eq:x}-\eqref{eq:H0}, the ergodic capacity is given by}
\begin{align}
\notag\bar{C}=&\max_{\mathbf{w},\mathbf{p}} B\log_2\left( 1+\left\langle\frac{||\mathbf{w}^{\textrm{H}}\mathbf{H}\mathbf{p}s||^2}{||\mathbf{w}^{\textrm{H}}\mathbf{n}||^2}\right\rangle\right)\\
\notag=&\max_{\mathbf{w},\mathbf{p}} B\log_2\left(
		1+\frac{P}{N_0B}\left\langle  ||\mathbf{w}^{\textrm{H}}\mathbf{H}\mathbf{p}||^2\right\rangle  \right)\\
\notag=&\max_{\{w_i,p_j\}}B\log_2\Bigg(1+\frac{P}{N_0B}\alpha_{\textrm{LoS}}^2\sum_{i,i'=1}^{N_t}\sum_{j,j'=1}^{N_r}p_ip_{i'}^{\textrm{H}}w_j^{\textrm{H}}\\
\notag&w_{j'}a_{ti}^{\textrm{H}}a_{ti'}a_{rj}a_{rj'}^{\textrm{H}}
		\left\langle \Psi_{ij}\Psi_{i'j'}^{*}
		\right\rangle\Bigg)\\
\notag=&\max_{\{w_i,p_j\}}B\log_2\Bigg(1+\frac{P}{N_0B}\alpha_{\textrm{LoS}}^2\sum_{i,i'=1}^{N_t}\sum_{j,j'=1}^{N_r}p_ip_{i'}^{\textrm{H}}w_j^{\textrm{H}}\\
\notag& w_{j'}a_{ti}^{\textrm{H}}a_{ti'}a_{rj}a_{rj'}^{\textrm{H}}
		\rho_{ij,i'j'}\sqrt{\left\langle \Psi_{ij}^2
			\right\rangle\left\langle \Psi_{i'j'}^2
			\right\rangle}\Bigg)\\
\notag=&\max_{\{w_i,p_j\}}B\log_2\Bigg(1+\frac{P}{N_0B}\alpha_{\textrm{LoS}}^2\alpha_{\textrm{turb}}^2\sum_{i,i'=1}^{N_t}\sum_{j,j'=1}^{N_r}p_i\\
\notag&p_{i'}^{\textrm{H}}w_j^{\textrm{H}}w_{j'}a_{ti}^{\textrm{H}}a_{ti'}a_{rj}a_{rj'}^{\textrm{H}}
		\rho_{iji'j'}\Bigg)\\
\le& B\log_2\left(1+\frac{P\alpha_{\textrm{LoS}}^2\alpha_{\textrm{turb}}^2}{N_0BN_tN_r}\sum_{i,i'=1}^{N_t}\sum_{j,j'=1}^{N_r}
\rho_{iji'j'}\right),
	\label{eq:C}
\end{align}
where the equation holds when $\mathbf{p}=\mathbf{a}_t(\theta_{t},\phi_{t})$ and $\mathbf{w}=\mathbf{a}_r(\theta_{r},\phi_{r})$.$\left\langle \Psi_{ij}^2\right\rangle=\left\langle \Psi_{i'j'}^2\right\rangle=\alpha_{\textrm{turb}}^2$ represents the additional turbulence-induced attenuation. 
\wjg{$a_{ti}$ and $a_{ri}$ represent the $i^{\textrm{th}}$ element of $\mathbf{a}_t$ and $\mathbf{a}_r$, respectively.}
$\rho_{iji'j'}$ represents the normalized covariance (NC) between $\Psi_{ij}$ and $\Psi_{i'j'}$ as
\begin{equation}
	\rho_{iji'j'}\triangleq\frac{\left\langle \Psi_{ij}\Psi_{i'j'}^{\wjg{*}}
		\right\rangle}{\sqrt{\left\langle \Psi_{ij}^2
			\right\rangle\left\langle \Psi_{i'j'}^2
			\right\rangle}}.
   \label{eq:rhoiji'j'}
\end{equation}
Parameter $\rho_{iji'j'}$ characterizes the NC between the SISO channel from the $i^{\textrm{th}}$ transmit antenna to the $j^{\textrm{th}}$ receive antenna and the channel from the $i'^{\textrm{th}}$ transmit antenna to the $j'^{\textrm{th}}$ receive antenna. For non-turbulent media,  $\rho_{iji'j'}=1$, and the ergodic capacity of the UM-MIMO system can be computed as $C(f,d)=B\log_2\left(1+P_{t}\alpha_{\textrm{LoS}}^2N_tN_r/N_0\right)$, which reveals that a $N_t\times N_r$ the UM-MIMO system achieves $N_tN_r$ array gain. While for turbulent media, $\rho_{iji'j'}\le 1$, and the array gain is given by 
\begin{equation}
	G_{\textrm{turb}}=\frac{1}{N_tN_r}\sum_{i,i'=1}^{N_t}\sum_{j,j'=1}^{N_r}
		\rho_{iji'j'}\le N_tN_r,
\end{equation}
which is less than the array gain for non-turbulent media.
The additional loss due to turbulent media by destroying the spatial coherence among transmit and receive antenna array is expressed as
\begin{equation}
L_{\textrm{LoSC}}
=10\log\frac{ N_t^2N_r^2}{\sum_{i,i'=1}^{N_t}\sum_{j,j'=1}^{N_r}
		\rho_{iji'j'}}.
\label{eq:LoSC1}
\end{equation}
\wjg{In the rest of this section, we derive a closed-form expression of $\rho_{iji'j'}$ based on Maxwell's equation in turbulent media, where the turbulent media is statistically characterized by the spatial power spectrum~\eqref{eq:Phi_Kolmo}.}
\subsection{Maxwell's Equation in Turbulent Media}
To further study the LoSC effect due to the refractive index fluctuation in turbulent media, we investigate the physical nature of EM wave propagation in turbulent media, i.e., the Maxwell's equation. By assuming a monochromatic EM wave in unbounded medium with varying refractive index, Maxwell's equations for the electric field vector $\mathbf{E}_i(\mathbf{r})$ emitted from the $i^{\textrm{th}}$ transmit antenna can be simplified as
\begin{equation}
\nabla^2\mathbf{E}_i(\mathbf{r})+k^2n^2(\mathbf{r})\mathbf{E}_i(\mathbf{r})+2\nabla[\mathbf{E}_i(\mathbf{r}) \nabla\log n(\mathbf{r})]=0,
\label{eq:E(R)}
\end{equation}
where $\mathbf{r}=(x,y,z)$ denotes a three-dimensional position vector. $\nabla^2=\partial^2/\partial x^2+\partial^2/\partial y^2+\partial^2/\partial z^2$ is the Laplacian operator.
By further considering that the variation of refractive index is small, i.e., $n_1({\mathbf{r}})\triangleq n({\mathbf{r}})-\langle n({\mathbf{r}}) \rangle\approx n({\mathbf{r}})-1$, we drop the last term of~\eqref{eq:E(R)} and express $n^2(\mathbf{r})\approx 1+2n_1(\mathbf{r})$.
Let $E_i(\mathbf{r})$ denote the scalar electric field as $E_i(\mathbf{r})=|\mathbf{E}_i(\mathbf{r})|$.
The equation~\eqref{eq:E(R)} can simplified as
\begin{equation}
    \nabla^2 E_i(\mathbf{r})+k^2E_i(\mathbf{r})+2k^2n_1(\mathbf{r})E_i(\mathbf{r})=0.
\label{eq:U(R)}
\end{equation}
Let $E_{i,0}(\mathbf{r})$ denote the solution of~\eqref{eq:U(R)} with $n_1(\mathbf{r})=0$, which represents the unperturbed electric field in non-turbulent media.
By replacing the location of the $j^{\textrm{th}}$ antenna in the receive array according to the coordinate represented by $(\mathbf{r}_j,L)$, the channel perturbation in~\eqref{eq:psi} can be expressed as 
\begin{equation}
\Psi_{ij}=E_{i}(\mathbf{r}_j,L)/E_{i,0}(\mathbf{r}_j,L).
\label{eq:psi_ij}
\end{equation}
\subsection{Rytov Approximation}
Due to the lack of a closed-form expression for the spatial distribution of refractive index $n_1(\mathbf{r}_j,L)$, it is 
difficult to solve~\eqref{eq:U(R)} directly. An approximated method called Rytov approximation is widely-used to solve~\eqref{eq:U(R)}, where the electric field is expanded as 
\begin{equation}
E_i(\mathbf{r}_j,L)=E_{i,0}(\mathbf{r}_j,L)\exp[\psi_{i,1}(\mathbf{r}_j,L)]\exp[\psi_{i,2}(\mathbf{r}_j,L)]\cdots,
\label{eq:Rytov}
\end{equation}
where $\psi_{i,q}(\mathbf{r}_j,L)$ denotes the $q^{\textrm{th}}$-order Rytov perturbation satisfying $|\psi_{i,q+1}(\mathbf{r}_j,L)|\ll |\psi_{i,q}(\mathbf{r}_j,L)|$.
Therefore, the covariance of turbulence-induced channel perturbation in~\eqref{eq:C} can be further expressed in~\eqref{eq:Gamma} at the bottom of \wjg{this page}.
\begin{figure*}[b]
	\hrulefill
\begin{equation}
	\begin{aligned}
		&\left\langle \Psi_{ij}\Psi_{i'j'}^*
		\right\rangle=\left\langle \exp[\psi_{i}(\mathbf{r}_j,L)+\psi_{i'}^*(\mathbf{r}_{j'},L)]
		\right\rangle\\
		=&\exp\Big[
		\langle \psi_{i}(\mathbf{r}_j,L)+\psi_{i'}^*(\mathbf{r}_{j'},L)
		\rangle
		+\frac{1}{2}(\langle [\psi_{i}(\mathbf{r}_j,L)
		+\psi_{i'}^*(\mathbf{r}_{j'},L)]^2
		\rangle-\langle [\psi_{i}(\mathbf{r}_j,L)+\psi_{i'}^*(\mathbf{r}_{j'},L)]
		\rangle^2)
		\Big]\\
            \overset{(a)}{\approx}&\exp\Big[
		\langle\psi_{i,1}(\mathbf{r}_j,L)\rangle
		+\langle\psi_{i,2}(\mathbf{r}_j,L)\rangle
		+\langle\psi_{i',1}(\mathbf{r}_{j'},L)\rangle+\langle\psi_{i',2}(\mathbf{r}_{j'},L)\rangle
		+\frac{1}{2}\Big(\langle\psi_{i,1}^2(\mathbf{r}_j,L)\rangle+\langle\psi_{i,1}(\mathbf{r}_j,L)\psi_{i',1}^{*}(\mathbf{r}_{j'},L)\rangle\\
		&+\langle\psi_{i,1}^{*}(\mathbf{r}_j,L)\cdot
		\psi_{i',1}(\mathbf{r}_{j'},L)\rangle+\langle\psi_{i',1}^2(\mathbf{r}_{j'},L)\rangle
		-\langle\psi_{i,1}(\mathbf{r}_j,L)\rangle^2-\langle\psi_{i,1}(\mathbf{r}_j,L)\rangle\langle\psi_{i',1}^{*}(\mathbf{r}_{j'},L)\rangle
		-\langle\psi_{i,1}^{*}(\mathbf{r}_j,L)\rangle\\
        &\langle\psi_{i',1}(\mathbf{r}_{j'},L)\rangle
		-\langle\psi_{i',1}(\mathbf{r}_{j'},L)\rangle^2
		\Big)
		\Big]\\
		=&\exp\Big[\langle \psi_{i,2}(\mathbf{r}_j,L)\rangle+\frac{1}{2}\langle \psi_{i,1}^2(\mathbf{r}_j,L)\rangle+\langle \psi_{i,2}(\mathbf{r}_{j'},L)\rangle+\frac{1}{2}\langle \psi_{i,1}^2(\mathbf{r}_{j'},L)\rangle+\langle \psi_{i,1}(\mathbf{r}_j,L)\psi_{i',1}^*(\mathbf{r}_{j'},L)\rangle\Big]\\
		=&\exp[M_{ij,1}(L)+M_{i'j',1}(L)+M_{ii'jj',2}(L)],
	\end{aligned}
	\label{eq:Gamma}
\end{equation}
\end{figure*}
The approximation (a) in~\eqref{eq:Gamma} omits all perturbations with third- and high-order terms.

According to~\eqref{eq:Gamma}, we note that the covariance of the turbulence-induced channel perturbation depends on two second-order moments denoted by 
$M_{ij,1}(L)$ and $M_{ii'jj',2}(L)$, which are respectively defined as
\begin{align}
	&M_{ij,1}(L)\triangleq\langle \psi_{i,2}(\mathbf{r}_j,L)\rangle+\frac{1}{2}\langle \psi_{i,1}^2(\mathbf{r}_j,L)\rangle,
	\label{eq:E1}
	\\
	&M_{ii'jj',2}(L)\triangleq\langle \psi_{i,1}(\mathbf{r}_j,L)\psi_{i',1}^*(\mathbf{r}_{j'},L)\rangle,
	\label{eq:E2}
\end{align}
where $M_{ij,1}(L)$ and $M_{ii'jj',2}(L)$ stand for the second-order statistical moment of first kind and second kind, respectively.

\subsection{Second-Order Moment Analysis of Turbulence-induced Perturbation}
Directly solving by Rytov approximation is still challenging because of the multiplication nature in~\eqref{eq:Rytov}.
We thereby expand the electric field in a linear expansion as 
\begin{equation}	
E_i(\mathbf{r}_j,L)=E_{i,0}(\mathbf{r}_j,L)+E_{i,1}(\mathbf{r}_j,L)+E_{i,2}(\mathbf{r}_j,L)\cdots,
\label{eq:linear}
\end{equation}
where $E_{i,q}(\mathbf{r}_j,L)$ represents the $q^{\textrm{th}}$-order linear perturbation of electric field at the $j^{\textrm{th}}$ Rx antenna satisfying $|E_{i,q+1}(\mathbf{r}_j,L)|\ll |E_{i,q}(\mathbf{r}_j,L)|$.
Then the first-order and second-order Rytov perturbation can be approximately expressed as 
\begin{align}
&\psi_{i,1}(\mathbf{r}_j,L)\approx \frac{E_{i,1}(\mathbf{r}_j,L)}{E_{i,0}(\mathbf{r}_j,L)},\\
&\psi_{i,2}(\mathbf{r}_j,L)\approx \frac{E_{i,2}(\mathbf{r}_j,L)}{E_{i,0}(\mathbf{r}_j,L)}-\frac{1}{2}\left[ \frac{E_{i,1}(\mathbf{r}_j,L)}{E_{i,0}(\mathbf{r}_j,L)}\right]^2.
\end{align}
By substituting the linear expansion~\eqref{eq:linear} into~\eqref{eq:U(R)}, the linear perturbation of the electric field can be expanded as a set of recursive equations as
\begin{align}
\nabla^2 E_{i,0}&(\mathbf{r}_j,L)+k^2E_{i,0}(\mathbf{r}_j,L)=0\\
\notag
\nabla^2 E_{i,q}&(\mathbf{r}_j,L)+k^2E_{i,q}(\mathbf{r}_j,L)=-2k^2n_1(\mathbf{r}_j,L)\\
&\quad\quad\quad\quad\quad\quad\quad\quad\quad\quad\quad\quad\cdot E_{i,q-1}(\mathbf{r}_j,L).
\label{eq:U(R)_recur}
\end{align}
By defining the normalized linear perturbation $\tilde{E}_{i,q}(\mathbf{r}_j,L)=E_{i,q}(\mathbf{r}_j,L)/E_{i,0}(\mathbf{r}_j,L)$, we can express the second-order moments as 
\begin{align}
	M_{ij,1}(L)&=\left\langle \tilde{E}_{i,2}(\mathbf{r}_j,L)\right\rangle,
	\label{eq:M1}
	\\
	M_{ii'jj',2}(L)&=\left\langle \tilde{E}_{i,1}(\mathbf{r}_j,L)\tilde{E}_{i',1}^*(\mathbf{r}_{j'},L)\right\rangle.
	\label{eq:M2}
\end{align}
To derive the second-order moments $M_{ij,1}(L)$ and $M_{ii'jj',2}(L)$, we begin from the analysis on the recursive differential equations~\eqref{eq:U(R)_recur}, which can be solved by the method of Green's function as
\begin{equation}
\begin{aligned}
E_{i,q}(\mathbf{r}_j,L)=\iiint_{\Omega} &G(\mathbf{r}_j,L,\mathbf{s},z)\cdot E_{i,q-1}(\mathbf{s},z)
\\
&\cdot 2k^2n_1(\mathbf{s},z)\d^2\mathbf{s}\d z,
\end{aligned}
	\label{eq:E_i,q}
\end{equation}
where $\Omega$ represents the entire scattering space. $\mathbf{s}$ and $z$ represent a vector and a scalar variable, respectively. The green's function $G(\mathbf{r}_j,L,\mathbf{s},z)$ is represented by
\begin{equation}
	G(\mathbf{r}_j,L,\mathbf{s},z)=\frac{1}{4\pi d(\mathbf{r}_j,L,\mathbf{s},z)}\exp\big[\textrm{i}kd(\mathbf{r}_j,L,\mathbf{s},z)\big],
\end{equation}
where $d(\mathbf{r}_j,L,\mathbf{s},z)=\sqrt{|\mathbf{r}_j-\mathbf{s}|^2+(L-z)^2}$.
For the UM-MIMO system, we apply far-field assumption where the distance between antennas in an array is much smaller than the propagation distance.
\begin{figure}
\centering
\includegraphics[width=0.48\textwidth]{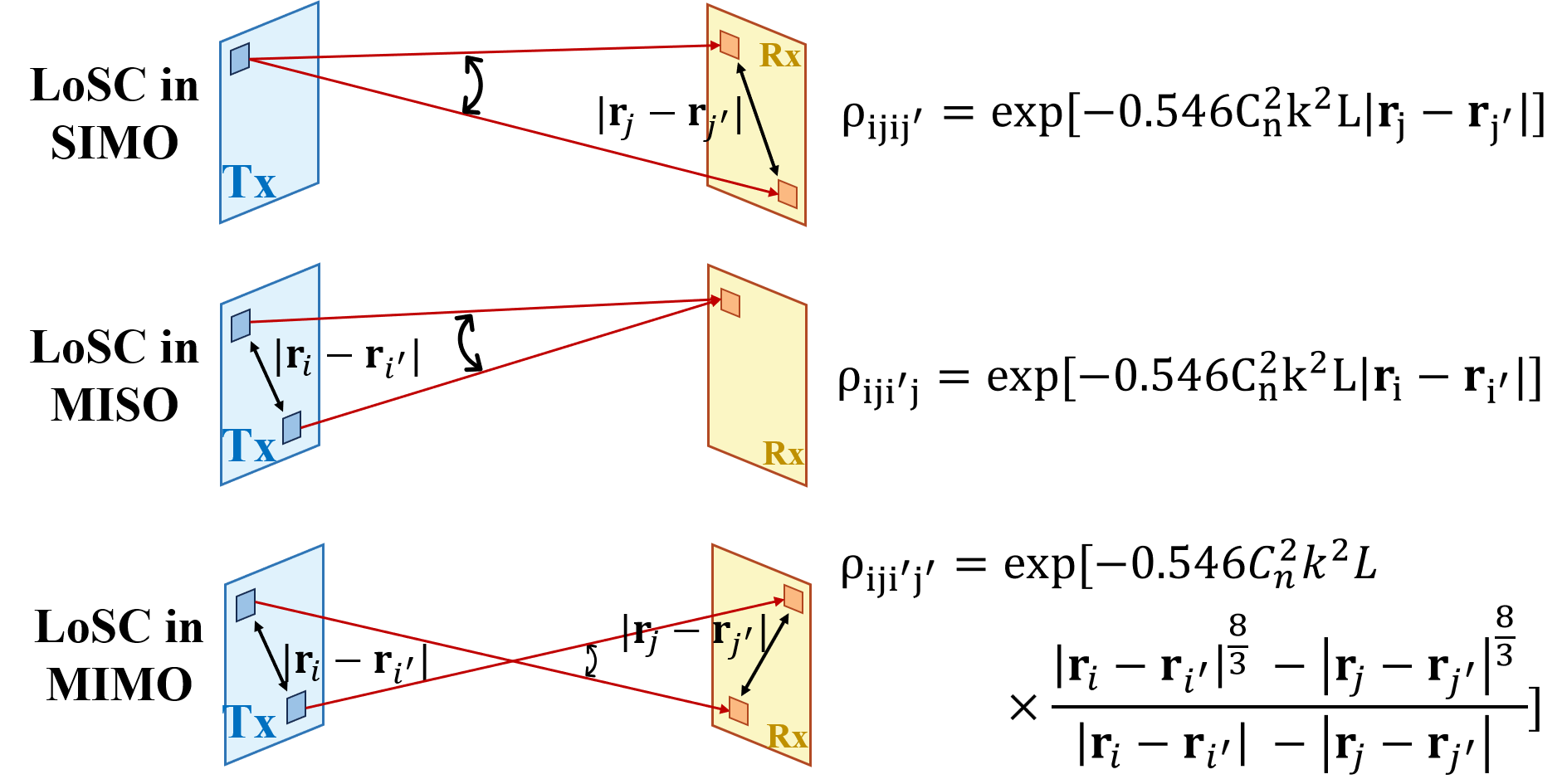}
		\captionsetup{font={footnotesize}}
    \caption{Illustration of LoSC effect for THz UM-MIMO system.}
	    \label{fig:LoSC_illu}
\end{figure}  
By applying the far-field approximation, the unperturbed electric field $E_{i,0}(\textbf{r}_j,L)$ emitted from the $i^{\textrm{th}}$ antenna located at $(\mathbf{r}_i,z_i)$ is expressed as
\begin{equation}
	\begin{aligned}
		&E_{i,0}(\textbf{r}_j,L)=\frac{\exp\left[\textrm{i}k\sqrt{(L-z_i+z_j)^2+(\mathbf{r}_i-\mathbf{r}_j)^2}\right]}{4\pi (L-z_i+z_j)}\\
		&\approx\frac{\exp\left[\textrm{i}kL+\frac{\textrm{i}kr_j^2}{2L}-\textrm{i}k(L-z_i+z_j)+\frac{\textrm{i}kr_i^2}{2L}-\frac{\textrm{i}k\mathbf{r}_i\cdot\mathbf{r}_j}{L}\right]}{4\pi L}.
	\end{aligned}
\label{eq:Uiq/Ui0}
\end{equation}
Then, by dividing both sides of~\eqref{eq:E_i,q} by $E_{i,0}(\mathbf{r}_j,L)$, we achieve the following inductive expression for $\tilde{E}_{i,q}(\mathbf{r}_j,L)$ as 
\begin{equation}
	\begin{aligned}
	&\tilde{E}_{i,q}(\mathbf{r}_j,L)=\frac{k^2}{2\pi}\int_{0}^{L}\d z\iint_{-\infty}^{\infty}\d^2s\exp\Bigg[\textrm{i}k(L-z)\\
	&+\frac{\textrm{i}k|\mathbf{s}-\mathbf{r}_j|^2}{2(L-z)}\Bigg] \frac{E_{i,0}(\mathbf{s},z)}{E_{i,0}(\mathbf{r}_j,L)}\tilde{E}_{i,q-1}(\mathbf{s},z)\frac{n_1(\mathbf{s},z)}{(L-z)},
	\end{aligned}
	\label{eq:Uiq/Ui0_L}
\end{equation}
where $\tilde{E}_{i,0}(\mathbf{r}_j,L)=1$.
We further express the refractive index fluctuation term $n_1(\mathbf{s},z)$ in the form of a two-dimensional Riemann-Stieltjes integral given by
\begin{equation}
	n_1(\textbf{s},z)=\iint_{-\infty}^{\infty}\exp(i\mathbf{k}\cdot\textbf{s})\d \upsilon(\mathbf{k},z),
	\label{eq:n1_integ}
\end{equation}
where $\d \upsilon(\mathbf{k},z)$ denotes the amplitude of refractive index and $\mathbf{k}=(\kappa_x,\kappa_y,0)$ represents the wave vector in rad/m.
By replacing~\eqref{eq:Uiq/Ui0} and~\eqref{eq:n1_integ} into~\eqref{eq:Uiq/Ui0_L}, we can deduce the following theorem about the normalized first-order linear perturbation of electric field.

\textbf{Theorem 1:}
\it The normalized first-order linear perturbation of electric field is given by
\rm
\begin{equation}
\begin{aligned}
	&\widetilde{E}_{i,1}(\mathbf{r}_j,L)=\textrm{i}k\int_{0}^{L}\d z\iint_{-\infty}^{\infty}\d \upsilon(\mathbf{k},z)\\
	&\cdot\exp\Big[-\frac{\textrm{i}z(L-z)}{2kL}\kappa^2-\frac{\textrm{i}z}{L}\mathbf{k}\cdot\textbf{r}_j+
	\frac{\textrm{i}(L-z)}{L}\mathbf{k}\cdot\mathbf{r}_i\Big].
\end{aligned}
	\label{eq:psi1}
\end{equation}
\it
where $\kappa=|\mathbf{k}|$.

Proof: 
\rm The detailed proof is provided in Appendix \ref{ap1}.
$\hfill\blacksquare$
\wjg{This theorem provides a closed-form expression for the first-order perturbation for the electric field in turbulent medium.}
By replacing~\eqref{eq:psi1} into~\eqref{eq:E1}, we can further express the first-kind second-order statistical moment of the perturbation term in the following theorem.

\textbf{Theorem 2:}
\it 
The first-kind second-order statistical moment $M_{ij,1}(L)$ can be expressed as 
\begin{equation}
	M_{ij,1}(L)=-2\pi^2k^2L\int_{0}^{1}\int_{0}^{\infty}\kappa \Phi_n(\kappa,L-L\xi)\d \kappa\d\xi,
	\label{eq:E1_res}
\end{equation}
where $\xi=1-z/L$. $\Phi_n(\kappa,z)$ is the Kolmogorov spectrum at position $z$.

Proof: 
\rm The detailed proof is provided in Appendix \ref{ap2}.
$\hfill\blacksquare$

Similarly, by replacing~\eqref{eq:psi1} into~\eqref{eq:E2}, we can further express the second-kind second-order statistical moment of the perturbation term in the following theorem.

\textbf{Theorem 3:} 
\it 
The second-kind second-order statistical moment of the perturbation term is expressed by
\rm
\begin{equation}
	\begin{aligned}
	M_{ii'jj',2}(L)&=4\pi k^2L\int_{0}^{1}\int_{0}^{\infty}\kappa\Phi_n(\mathbf{\kappa},L-L\xi)J_0(\kappa|(1-\xi)\\
	&(\mathbf{r}_j-\mathbf{r}_{j'})+\xi(\mathbf{r}_i-\mathbf{r}_{i'})|)\d\kappa \d\xi.
	\end{aligned}
	\label{eq:E2_res}
\end{equation}
\it
where $\Phi_n(\kappa,z)$ is the Kolmogorov spectrum at position $z$.

Proof: 
\rm The detailed proof is provided in Appendix \ref{ap3}.
$\hfill\blacksquare$

By replacing the results~\eqref{eq:E1_res} and~\eqref{eq:E2_res} into~\eqref{eq:Gamma}, the covariance of the turbulence-induced perturbation term is given by
\begin{equation}
	\begin{aligned}
		&\left\langle\Psi_{ij}\Psi_{i'j'}^{*}\right\rangle=
		\exp
		\Bigg\{
		-4\pi^2k^2L
		\int_{0}^{1}
		\int_{0}^{\infty}
		\kappa\Phi_n(\kappa,L-L\xi)
		\\
		&\cdot\bigg\{1-J_0\big[\kappa\xi\mathbf{r}_{jj'}+\kappa(1-\xi)\mathbf{r}_{ii'}\big]\bigg\}
		\d\kappa\d\xi
		\Bigg\},
	\end{aligned}
\label{eq:Psi_res}
\end{equation}
where $ \mathbf{r}_{ii'}=|\mathbf{r}_i-\mathbf{r}_{i'}|$ and $ \mathbf{r}_{jj'}=|\mathbf{r}_j-\mathbf{r}_{j'}|$.
\wjg{Theorem 2 and 3 present second-order statistics of the turbulence-induced perturbation, where $M_{ij,1}$ indicates the mean of second-order electric field perturbation while $M_{ii'jj',2}(L)$. represents the correlation between two first-order perturbation terms.}

\subsection{Additional Attenuation due to Loss of Spatial Coherence}
Further closed-form expression of~\eqref{eq:Psi_res} does not exist due to the non-uniform $\Psi_n(\kappa,z)$. For further simplification we assume that the RISC is uniform along the path, which is valid for horizontal or short-distance UAV links, i.e., $\Psi_n(\kappa,z)=\Psi_n(\kappa)$. Then, we can further express $\rho_{iji'j'}$ according to the following theorem.

\textbf{Theorem 4:} 
\it 
The NC between the turbulence-induced channel perturbation from the $i^{\textrm{th}}$ transmit antenna to the the $j^{\textrm{th}}$ receive antenna and from the $i'^{\textrm{th}}$ transmit antenna to the the $j'^{\textrm{th}}$ receive antenna for constant-RISC medium is given by 
\rm
\begin{equation}
	\rho_{iji'j'}=\exp\Bigg(-0.546C_n^2k^2L\frac{\sum_{p=0}^{7}\mathbf{r}_{ii'}^{\frac{p}{3}}\mathbf{r}_{jj'}^{\frac{7-p}{3}}}{\sum_{p=0}^{2}\mathbf{r}_{ii'}^{\frac{p}{3}}\mathbf{r}_{jj'}^{\frac{2-p}{3}}}\Bigg),
\end{equation}
\it 
which can alternatively expressed by
\rm
\begin{equation}
	\rho_{iji'j'}=
\begin{cases}
\exp\Big(-0.546C_n^2k^2L\frac{\mathbf{r}_{ii'}^{\frac{8}{3}}-\mathbf{r}_{jj'}^{\frac{8}{3}}}{\mathbf{r}_{ii'}-\mathbf{r}_{jj'}}\Big),
&\mathbf{r}_{ii'}\neq\mathbf{r}_{jj'},\\
\exp\left(-1.457C_n^2k^2L\mathbf{r}_{ii'}^{\frac{5}{3}}\right),
	&\mathbf{r}_{ii'}=\mathbf{r}_{jj'}.
\end{cases}
\label{eq:rho_case}
\end{equation}
\it
Proof: 
\rm The detailed proof is provided in Appendix \ref{ap4}.
$\hfill\blacksquare$

\wjg{Theorem 4 demonstrates that the atmospheric turbulence may cause partially-dependent or even independent amplitude and phase shifts to different SISO channels in a MIMO system. The correlation of the turbulence-induced perturbation, characterized by $\rho_{iji'j'}$ depends on how far apart their transmit antennas and receive antennas are from each other.
}
The NC for different SISO channels in the UM-MIMO are summarized and depicted in Fig.~\ref{fig:LoSC_illu}, including the single-input multiple-output (SIMO), multiple-input single-output (MISO), and MIMO cases. The expressions of NC for the SIMO and MISO case match with the the spatial coherence of optical communication system. The LoSC in MIMO system presents a general expression for the channel covariance between antenna pairs in turbulent media.
By substituting~\eqref{eq:rho_case} into~\eqref{eq:LoSC1}, the additional loss due to LoSC is finally expressed in~\eqref{eq:LoSC2} at the top of \wjg{this page}.
\begin{figure*}[!t]
\begin{equation}
L_{\textrm{LoSC}}
=10\log\frac{ N_t^2N_r^2}{\sum_{i,i'=1}^{N_t}\sum_{j,j'=1}^{N_r}
\exp\left(-0.546C_n^2k^2L\frac{\sum_{p=0}^{7}\mathbf{r}_{ii'}^{\frac{p}{3}}\mathbf{r}_{jj'}^{\frac{7-p}{3}}}{\sum_{p=0}^{2}\mathbf{r}_{ii'}^{\frac{p}{3}}\mathbf{r}_{jj'}^{\frac{2-p}{3}}}\right),
}.
\label{eq:LoSC2}
\end{equation}    
	\hrulefill
\end{figure*}

\section{Fading and Attenuation Modeling of Turbulence in the Terahertz Band}~\label{sec:fading_and_attenuation}
In this section, we investigate the fading and attenuation led by turbulent media in the THz band. The turbulence-induced fading stands for the random power fluctuation of the received signal, and the turbulence-induced attenuation represents an additional attenuation besides the free-space path loss, molecular absorption, and scattering.  
\subsection{Turbulence-induced Fading in the Terahertz Band}~\label{sec:fading}
Unlike traditional multi-path fading led by random constructive or destructive interference of multi-path signals, turbulence-induced is due to the random direction distortion of the EM wave. 
%By assuming the received signal power as $P_r$, we define the fading parameter $I$ as the instantaneous ratio of the signal intensity to its statistical average, i.e., $I={P_r}/{\langle P_r\rangle}$. 
Previous studies have proposed several statistical models to characterize the turbulence-induced fading~\cite{dordova2010calculation} corresponding to the different strengths of turbulence, including the log-normal for weak turbulence, the K distribution for strong turbulence, and the exponential distribution corresponding to the saturation regime~\cite{dordova2010calculation}.
The strength of turbulence can be distinguished according to the Rytov variance, given by 
\begin{equation}
	\sigma_R^2=0.5C_n^2k^{7/6}L^{11/6},
\end{equation}
The three conditions $\sigma_R^2\ll 1$, $\sigma_R^2\sim 1$, and $\sigma_R^2\gg 1$ correspond to the case of weak turbulence, strong turbulence, and saturated regime, respectively.  
In extreme cases, e.g., when the propagation distance is $100\,\textrm{km}$ and the RISC is $C_n^2=10^{-11}\,\textrm{m}^{-2/3}$, the Rytov variance at $300\,\textrm{GHz}$ is as large as $\sigma_{R}^2=396$. Therefore, the strength of turbulence can fall in all three cases for THz UAV communications.  
A universal model applicable under weak turbulence, strong turbulence or saturated regime conditions is the Gamma-Gamma distribution~\cite{al2001mathematical}. By applying Kolmogorov's theory, the Gamma-Gamma model assumes that the fluctuation of turbulence is caused by large-scale eddies and small-scale eddies, i.e., $\Psi=\Psi_a\Psi_b$, and the two terms are governed by independent Gamma distributions, given by
\begin{align}
	p_{\Psi_a}(\Psi_a)=&\frac{\alpha_c(\alpha_c \Psi_a)^{\alpha_c-1}}{\Gamma(\alpha_c)}\exp(-\alpha_c \Psi_a), \Psi_a>0, \alpha_c>0,\\
	p_{\Psi_b}(\Psi_b)=&\frac{\beta_c(\beta_c \Psi_a)^{\beta_c-1}}{\Gamma(\beta_c)}\exp(-\beta_c \Psi_b), \Psi_b>0, \beta_c>0,
\end{align}
where $\alpha_c$ represents the effective number of large-scale cells, and $\beta_c$ represents the effective number of small-scale ones.
According to the chain rule and the independency of $\Psi_a$ and $\Psi_b$, the fading parameter $\Psi$ follows a Gamma-Gamma distribution, which is expressed by 
\begin{equation}
	p_\Psi(\Psi)=\frac{2(\alpha_c\beta_c)^{\frac{\alpha_c+\beta_c}{2}}}{\Gamma(\alpha_c)\Gamma(\beta_c)}\Psi^{\frac{\alpha_c+\beta_c}{2}-1}K_{\alpha_c-\beta_c}\big[\sqrt{2(\alpha_c\beta_c \Psi)}\big],
	\label{eq:GammaGamma}
\end{equation}
where $\Gamma(\cdot)$ denotes the Gamma function. $K_{p}(\cdot)$ is the modified Bessel function of the second kind of $p^{\textrm{th}}$ order. 
The large-scale and small-scale fading parameters $\alpha_c$ and $\beta_c$ can be modeled according to Andrew's method~\cite{al2001mathematical}, which are expressed as
\begin{align}
	\alpha_c=&\left[\exp\left(\frac{0.49\sigma_R^2}{(1+0.18D_{ra}^2+0.56\sigma_R^{12/5})^{7/6}}\right)-1\right]^{-1},
	\label{eq:alpha}
	\\
	\beta_c=&\left[\exp\left(\frac{0.51\sigma_R^2(1+0.69D_{ra}^2\sigma_R^{12/5})^{-5/6}}{(1+0.9D_{ra}^2+0.62\sigma_R^{12/5})^{7/6}}\right)-1\right]^{-1},
	\label{eq:beta}
\end{align}
where $D_{ra}=\sqrt{kl_{ra}^2/4L}$, and $l_{ra}$ represents the diameter of the antenna receiving aperture. Given that the effective area of the received antenna is $A_{\textrm{eff}}=\lambda^2/4\pi$, we have $l_{ra}=\lambda/\pi$. The four statistical models for turbulence fading are summarized in Table.~\ref{table:fading}. The relationship between the Gamma-Gamma distribution with the log-normal model, K distribution, and exponential distribution are elaborated as follows.
\begin{itemize}
	\item     When $\sigma_R^2<1$, i.e., in the case of weak turbulence, we have $\alpha_c\gg 1$ and $\beta_c\gg 1$ and the Gamma-Gamma distribution is approximately a log-normal distribution. 
	\item     When $\sigma_R^2>1$, i.e., in the case of strong turbulence, we have $\beta_c\approx 1$, and the Gamma-Gamma distribution shrinks to a K distribution. 
	\item     When $\sigma_R^2\to \infty$, i.e., in the saturated regime, we have $\alpha_c\gg 1$ and $\beta_c\approx 1$, and the Gamma-Gamma distribution approximately follows an exponential distribution. 
\end{itemize}
\begin{table*}[ht]  
	\caption{Statistical models of turbulence-induced fading.} 
	\label{table:fading}
	\centering
	\renewcommand\arraystretch{1}
	\begin{tabular}{p{5cm}p{5cm}p{7cm}} 
		\hline  
		\hline  
		\textbf{Name} & \textbf{Condition} & \textbf{Expression}\\ 
		Log-normal & $\sigma_R^2\ll 1$, Weak turbulence & $p(\Psi)=\frac{1}{\sqrt{2\pi\sigma_{\Psi}^2}\Psi}\exp\left[-{(\ln \Psi)^2}/{2\sigma_\Psi^2}\right]$\\
		K distribution & $\sigma_R^2> 1$, Strong turbulence &
		$p(\Psi)=\frac{2\alpha_c}{\Gamma(\alpha_c)}(\alpha_c \Psi)^{(\alpha_c-1)/2}K_{\alpha_c-1}(2\sqrt{\alpha_c \Psi})$
		\\
		Exponential distribution & $\sigma_R^2\gg 1$, Saturated regime & $p(\Psi)=\frac{1}{b}\exp(-\Psi/b)$
		\\
		Gamma-Gamma distribution & All & Equation~\eqref{eq:GammaGamma}
		\\
		\hline
		\hline  
	\end{tabular}  
\end{table*} 

\subsection{Attenuation Effect of Turbulence in the Terahertz Band}
Similar to the fading model of turbulence, the attenuation model lacks deterministic and closed-form solutions due to the complexity and difficulty of solving the Navier-Stokes equations. An empirical formula developed by Larry C. Andrews for the turbulent attenuation $L_{\textrm{tur}}$ can be expressed as
\begin{equation}
	L_{\textrm{tur}} =10\log\left|1-\sqrt{\sigma_{\Psi}^2}\right|,
\end{equation}
where $\sigma_{\Psi}^2\triangleq\langle \Psi^2\rangle$ denotes the variance of $\Psi$. Given the Gamma-Gamma distribution of the $\Psi$, the variance of $\Psi$ is expressed as
\begin{equation}
	\sigma_\Psi^2
	=\langle \Psi_a^2\rangle \langle \Psi_b^2\rangle =\frac{1}{\alpha_c}+\frac{1}{\beta_c}+\frac{1}{\alpha_c\beta_c}.        
	\label{eq:atten}
\end{equation}
By substituting~\eqref{eq:alpha} and~\eqref{eq:beta}
into~\eqref{eq:atten}, the attenuation caused by turbulence in the THz band can be expressed in~\eqref{eq:Ltur} at the bottom of \wjg{this page}.
\setcounter{TempEqCnt}{\value{equation}}
\begin{figure*}[!b]
	\hrulefill
	\begin{equation}
		\begin{aligned}
			L_{\textrm{tur}}=&10\log_{10}\Bigg|1-\textrm{sqrt}\Bigg\{
			\exp\left(\frac{0.49\sigma_R^2}{(1+0.18D_{ra}^2+0.56\sigma_R^{12/5})^{7/6}}\right)
			+\exp\left(\frac{0.51\sigma_R^2(1+0.69D_{ra}^2\sigma_R^{12/5})^{-5/6}}{(1+0.9D_{ra}^2+0.62\sigma_R^{12/5})^{7/6}}\right)-2\\
			&+\left[\exp\left(\frac{0.49\sigma_R^2}{(1+0.18D_{ra}^2+0.56\sigma_R^{12/5})^{7/6}}\right)-1\right]\times\left[\exp\left(\frac{0.51\sigma_R^2(1+0.69D_{ra}^2\sigma_R^{12/5})^{-5/6}}{(1+0.9D_{ra}^2+0.62\sigma_R^{12/5})^{7/6}}\right)-1\right]
			\Bigg\}\Bigg|.
		\end{aligned}
		\label{eq:Ltur}
	\end{equation}
\end{figure*}

\begin{figure}[h!]
\centering
 \includegraphics[width=0.45\textwidth]{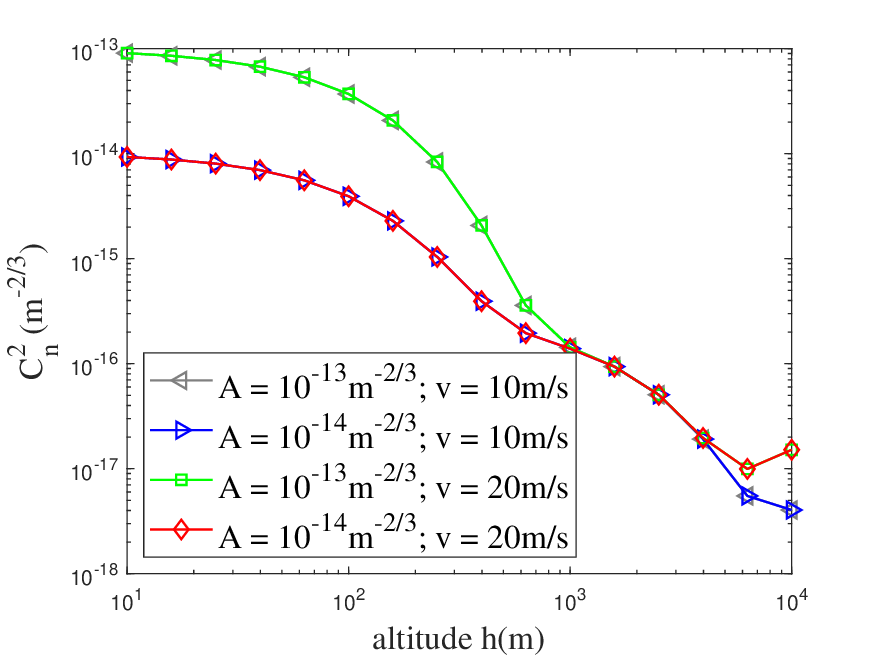}
	\captionsetup{font={footnotesize}}
	\caption{RISC at the different altitudes. $A$ represents the terrestrial RISC. $v$ stands for the average wind velocity.}
	\label{fig:RISC}
\end{figure}
\begin{figure}[t]
		\centering
		\subfigure[]{
			\includegraphics[width=0.45\textwidth]{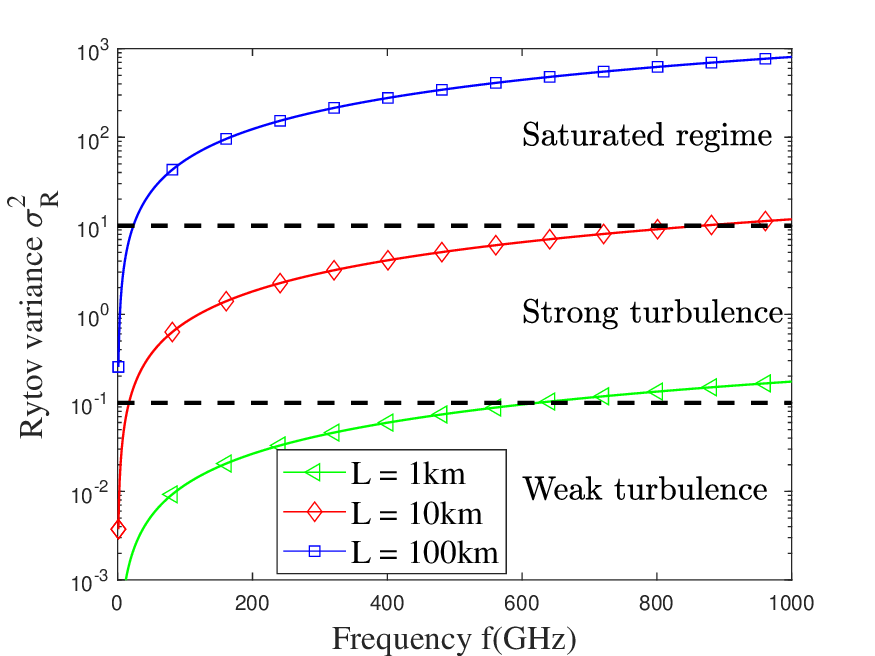}
			\label{fig:fig_rv_f_L}
		}
		\subfigure[]{\includegraphics[width=0.45\textwidth]{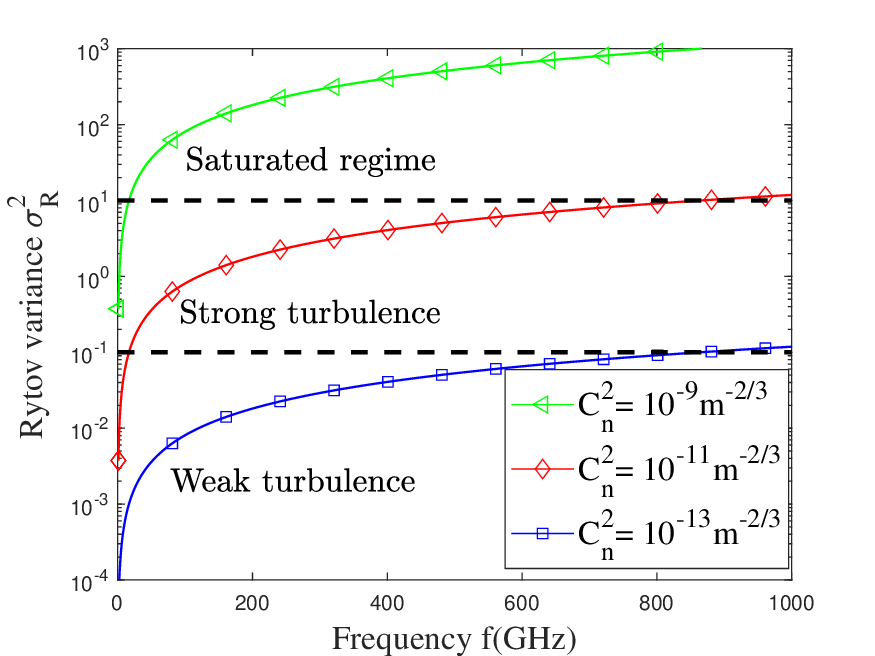}
			\label{fig:fig_rv_f_cn2}
		}
		\captionsetup{font={footnotesize}}
		\caption{Rytov variance with the different frequency. (a) With varying propagation distance $L$; (b) With varying RISC $C_n^2$. }
		\label{fig:rv_f}
	\end{figure} 

    \section{Numerical Results}~\label{sec:numerical}
    In this section, we perform numerical evaluations on the attenuation and LoSC caused by atmospheric turbulence on THz UM-MIMO UAV wireless communications.
    Specifically, the RISC and the Rytov variance, as two critical parameters to characterize the strength of the turbulence, are first investigated. The altitude dependency of the RISC in the THz band and the distance-dependent Rytov variance are illustrated.
    Then, the NCs among different Tx and Rx antenna pairs in MIMO are computed by varying the propagation distance. The additional attenuation due to the LoSC is computed under the different RISC, array size, and antenna spacing.
    Furthermore, the fading and attenuation caused by the turbulence are analyzed to illustrate the influence of turbulent media on THz UAV channel. Unless specified, the parameters in the simulations are listed in Table~\ref{Table}.
	\begin{table} 
		\caption{Simulation Parameters}
		\label{Table}
		\begin{tabular}{p{1cm}p{4cm}p{0.8cm}p{1.2cm}} 
			\hline  
			\hline  
			\textbf{Notation} & \textbf{Definition} & \textbf{Value} & \textbf{Unit}\\  
			\hline 
			$P_{t}$ & Transmit power & 10 & dBm\\
			$N_{tx}$ & \# of row antennas at Tx& 32 & -\\
			$N_{ty}$ & \# of column antennas at Tx& 32 & -\\			
			$N_{\textrm{rx}}$ & \# of row antennas at Rx& 32 & -\\ 
			
			$N_{\textrm{ry}}$ & \# of column antennas at Rx & 32 &-\\ 			
			$N_0$ & AWGN PSD & -174 & dBm/Hz\\
			
			$f$ & Carrier frequency & 300 & GHz\\
			\wjg{$B$} & \wjg{Bandwidth} & \wjg{1} & \wjg{GHz}\\
			$L$ & Propagation distance & 1 & km\\
			$C_n^2$ & RISC & $10^{-9}$ & $\textrm{m}^{-2/3}$\\
			\hline
			\hline  
		\end{tabular}  
	\end{table}

    \subsection{Refractive Index Structure Constant and Rytov Variance}
 The RISC at the different altitudes modeled \wjg{based on~\eqref{eq:RISC}-\eqref{eq:cn2_transform}} is shown in Fig. 4 with varying terrestrial RISC $A$ and average wind speed $v$. 
    We observe that the RISC shows a decreasing trend as the altitude increases.
    In the low-altitude regions ($h<1\,\textrm{km}$), the terrestrial RISC $A$ primarily governs the RISC value\wjg{ and the difference due to wind speed is small}, while in the high-altitude region ($h>1\,\textrm{km}$), the wind speed dominates the trend of RISC\wjg{ and the terrestrial RISC $A$ has weak effects on the curve.}

The Rytov variance determines the strength of the attenuation effect of atmospheric turbulence on THz wave propagation.
The Rytov variances under different frequencies in the THz band are shown in Fig.~\ref{fig:rv_f}. 
In Fig.~\ref{fig:fig_rv_f_L} for varying propagation distance $L$ and RISC $C_n^2$, the RISC is $C_n^2=10^{-11}\,\textrm{m}^{-2/3}$, and in Fig.~\ref{fig:fig_rv_f_cn2}, the propagation distance is $L=10\,\textrm{km}$. 
We approximately regulate the range of weak turbulence, strong turbulence, and saturated regime as $\sigma_R^2<0.1$, $0.1\le\sigma_R^2\le 10$, and $\sigma_R^2>10$, respectively.

    \subsection{Loss of Spatial Coherence} 
By using~\eqref{eq:rhoiji'j'}, the LoSC is investigated by analyzing the NC between two SISO channels.
In Fig.~\ref{fig:rho} the NC values for the different distances between transmit antennas and receive antennas are investigated. 
We observe that $\rho_{iji'j'}$ decreases as $|\mathbf{r}_i-\mathbf{r}_{i'}|$ and $|\mathbf{r}_i-\mathbf{r}_{i'}|$ increase. Moreover, as the propagation distance $L$ increases from $1\,\textrm{km}$ to $100\,\textrm{km}$, the decreasing trend of $\rho_{iji'j'}$ is faster. This implies that as the propagation distance increases, the influence of atmospheric turbulence is severer, and the resulting LoSC is therefore more significant.

	\begin{figure*}[htbp]
		\centering
		\subfigure[]{
			\includegraphics[width=0.3\textwidth]{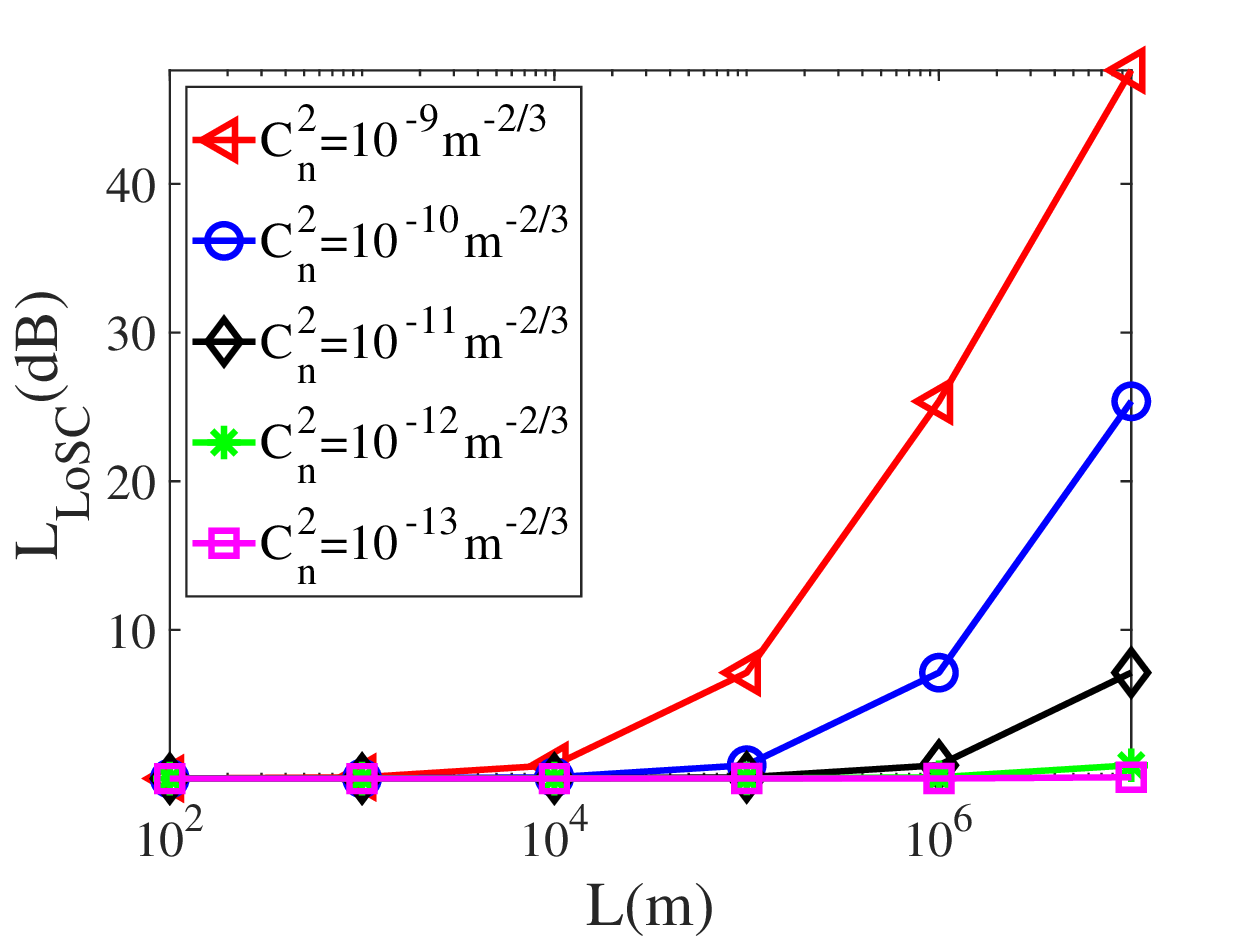}
			\label{fig:L_LoSC_L_cn2}
		}
		\subfigure[]{
			\includegraphics[width=0.3\textwidth]{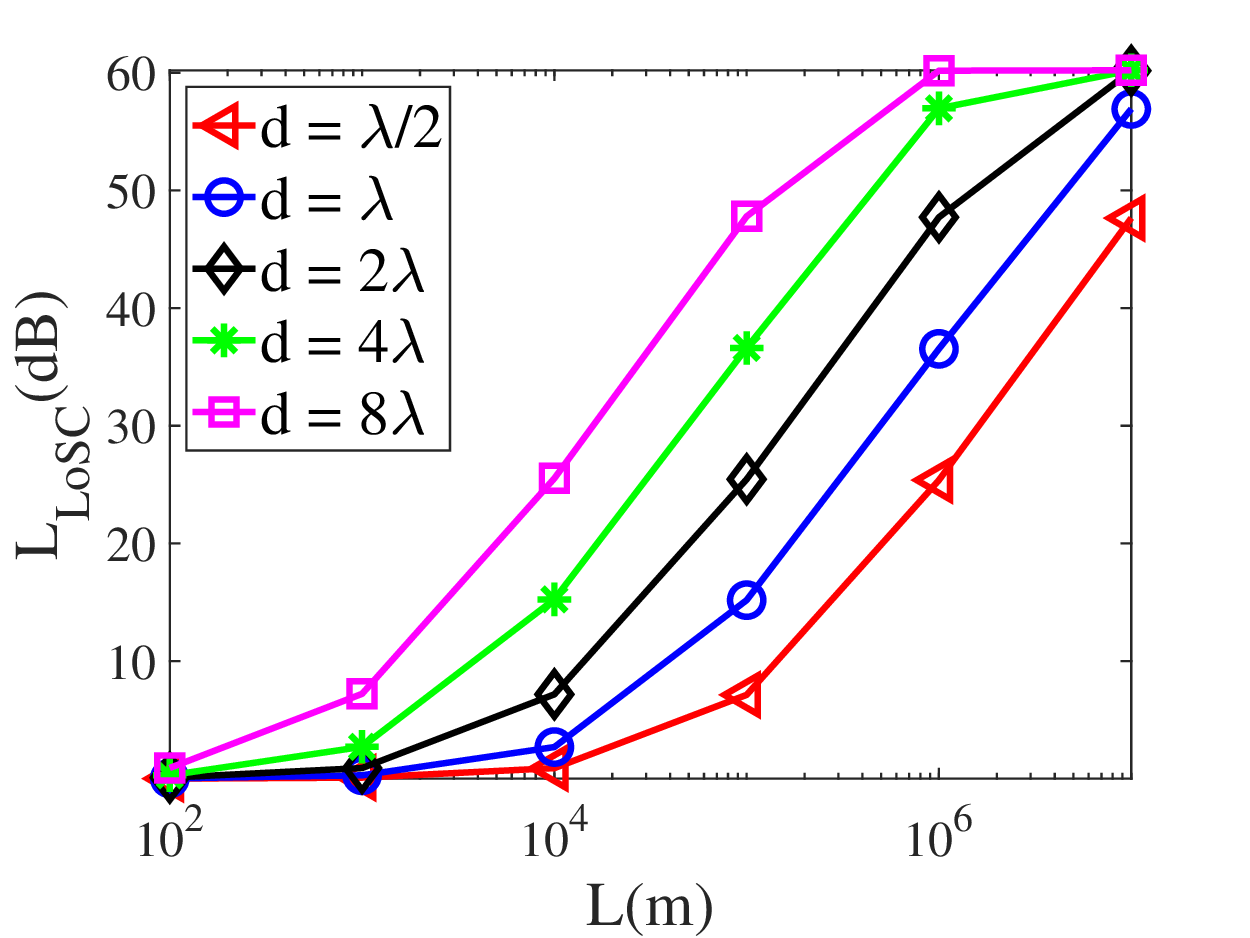}
			\label{fig:L_LoSC_L_d}
		}
            \subfigure[]{
			\includegraphics[width=0.3\textwidth]{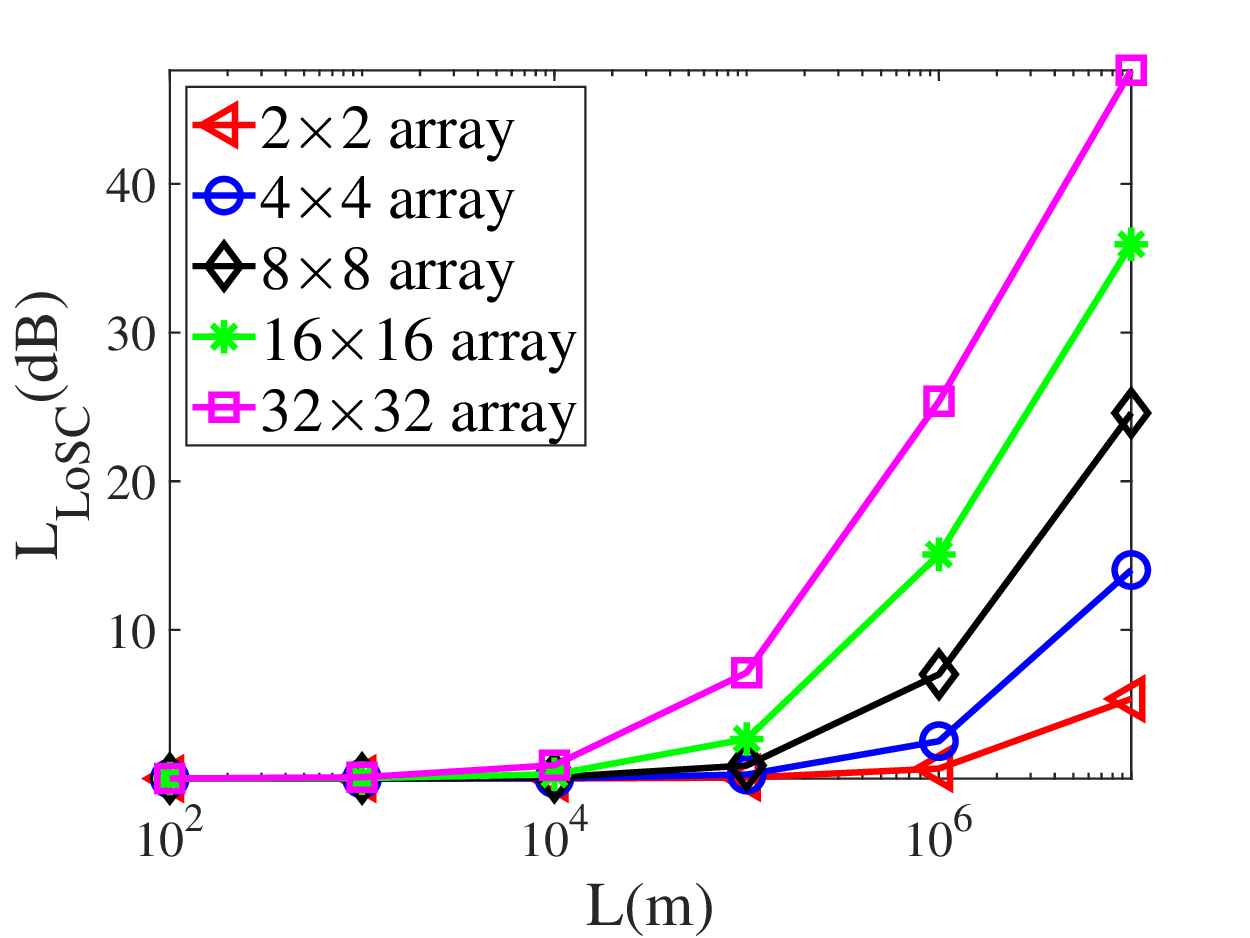}
			\label{fig:L_LoSC_L_Nt}
		}
		\captionsetup{font={footnotesize}}
		\caption{Attenuation due to LoSC versus distance. (a) With different RISC $C_n^2$; (b) With different antenna spacing $d$; (c) With different array size.}
		\label{fig:L_LoSC_L}
		\captionsetup{font={footnotesize}}
	\end{figure*} 
\begin{figure}[!htbp]
    \centering
    \includegraphics[width=0.45\textwidth]{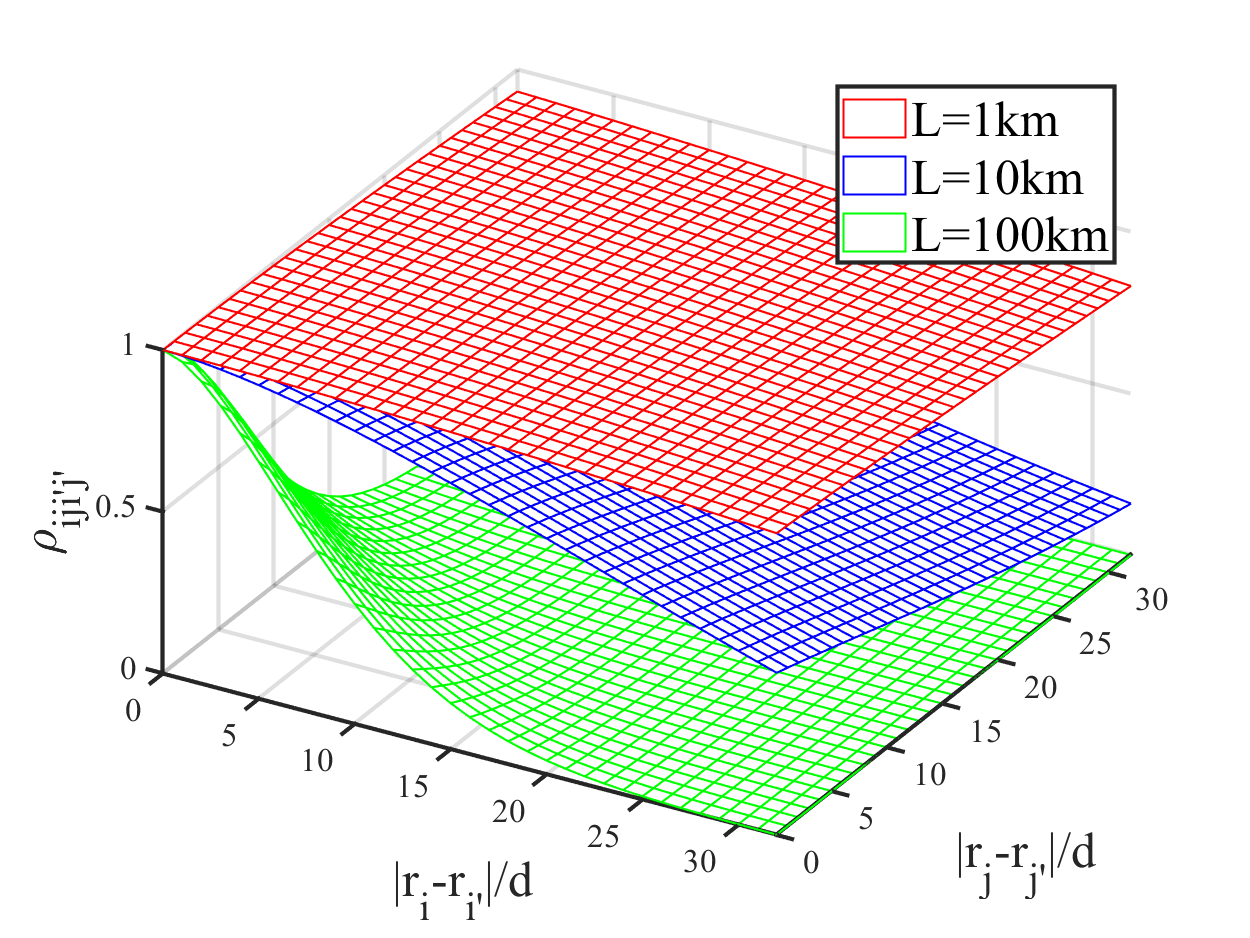}
    \captionsetup{font={footnotesize}}
    \caption{Normalized covariance between two SISO channel $\rho_{iji'j'}$ versus different $|\mathbf{r}_i-\mathbf{r}_{i'}|/d$ and $|\mathbf{r}_j-\mathbf{r}_{j'}|/d$ with varying propagation distances.}
    \label{fig:rho}
\end{figure}
We analyze the additional attenuation due to the LoSC with the different parameters. By choosing the frequency $f=300\,\textrm{GHz}$, the attenuation due to LoSC with the different propagation distances is shown in Fig.~\ref{fig:L_LoSC_L_cn2} with the different RISCs, in Fig.~\ref{fig:L_LoSC_L_Nt} for the different antenna size, and Fig.~\ref{fig:L_LoSC_L_d} for the different antenna spacing, respectively. The attenuation due to LoSC increases with higher RISC, larger antenna array, and wider antenna spacing.

\subsection{Fading and Attenuation Caused by Atmospheric Turbulence}

\begin{figure}[htbp]
		\centering
        \includegraphics[width=0.45\textwidth]{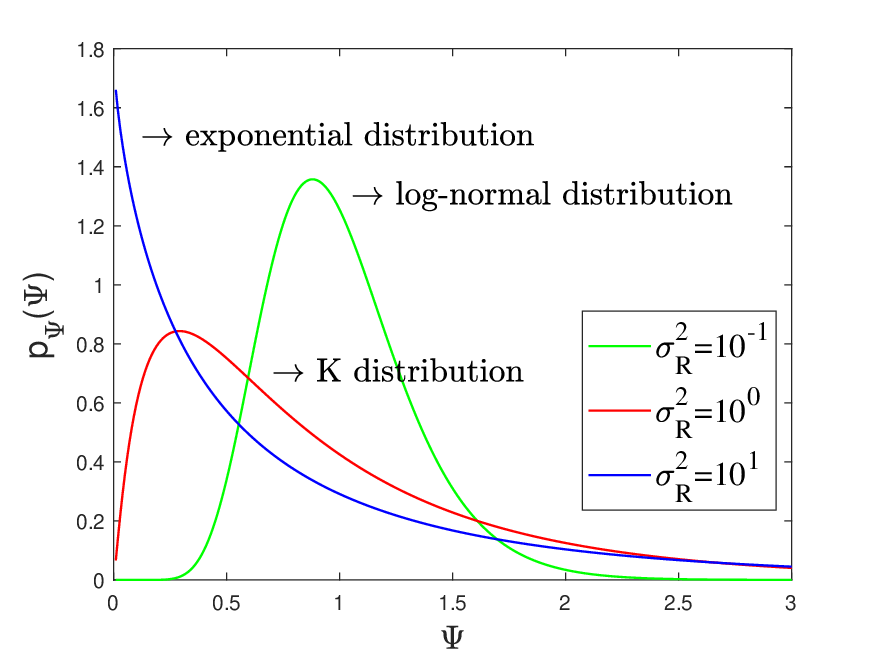}
		\captionsetup{font={footnotesize}}
        \caption{Gamma-Gamma distributed PDF for the different Rytov variances.}
	    \label{fig:GG_rv}
	\end{figure}

\begin{figure}[htbp]
		\centering
		\subfigure[]{
			\includegraphics[width=0.45\textwidth]{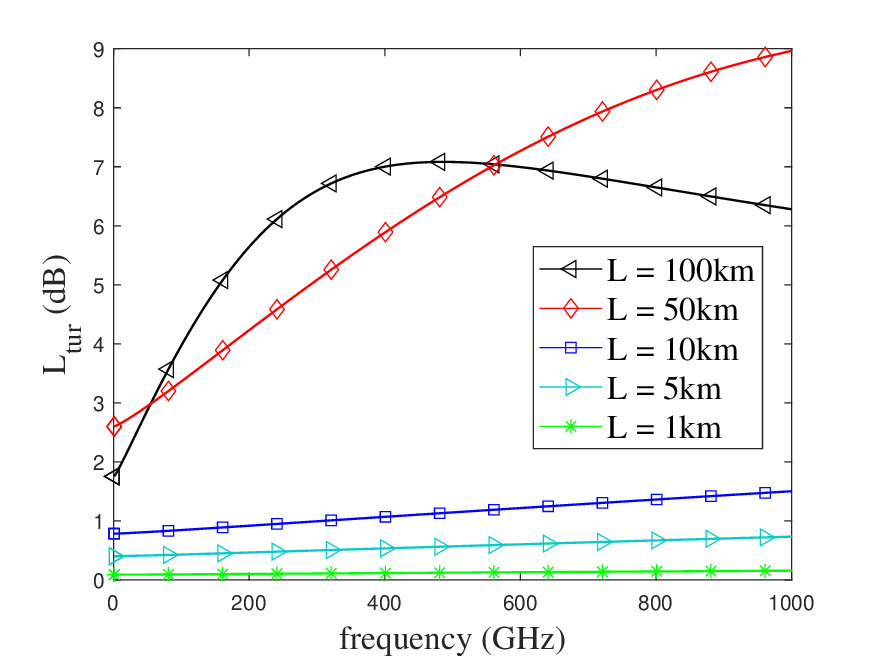}
			\label{fig:Ltur_f_L}
		}
		\subfigure[]{
			\includegraphics[width=0.45\textwidth]{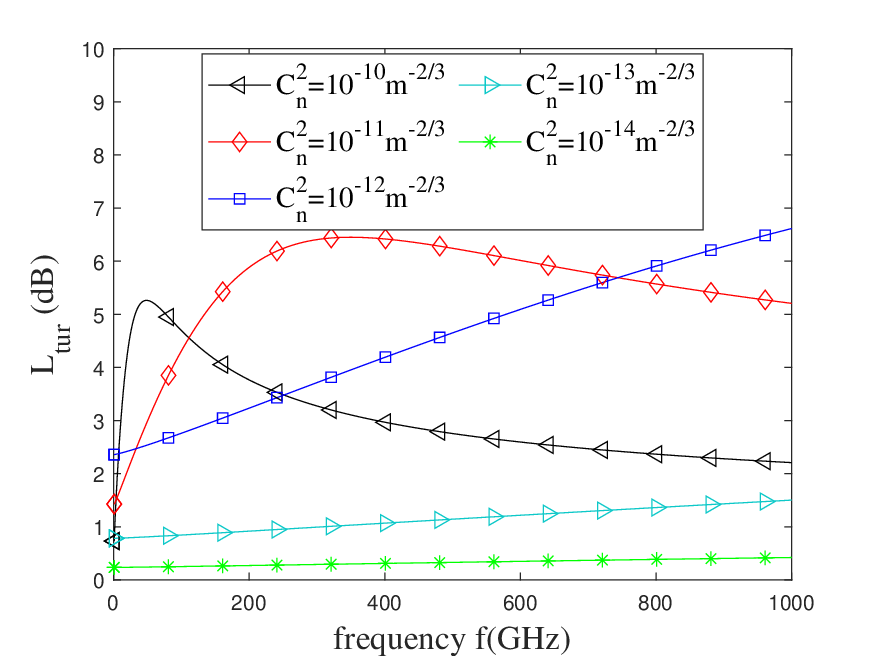}
			\label{fig:Ltur_f_cn2}
		}
		\captionsetup{font={footnotesize}}
		\caption{Attenuation caused by atmospheric turbulence with different frequencies. (a) With different propagation distance $L$; (b) With different RISC $C_n^2$. }
		\label{fig:Ltur_f}
		\captionsetup{font={footnotesize}}
	\end{figure}

The probability density function (PDF) of the Gamma-Gamma-distributed turbulence fading is shown in Fig.~\ref{fig:GG_rv}. 
The varying Rytov variance are taken as 0.1, 1, and 10, which corresponds to the weak turbulence, strong turbulence, and saturated regime, respectively. 
As we can observe, the PDFs for the three cases approximately follow log-normal, K distribution, and exponential distribution. 
For weak turbulence where $\sigma_R^2=0.1$, we have $\alpha_c=20.76$ and $\beta_c=19.75$, which indicates that both the numbers of effective large-scale and small-scale cells are large, and it approximates a log-normal distribution due to the law of large number. 
When $\sigma_R^2=1$, the turbulence is strong and we have $\alpha_c=2.95$ and $\beta_c=2.46$. 
For the saturated regime where $\sigma_R^2=10$, we have $\alpha_c=2.48$ and $\beta_c=0.98$. 
The attenuation caused by turbulence versus frequency with varying propagation distance $L$ and RISC $C_n^2$ is shown in Fig.~\ref{fig:Ltur_f}. 
As $L$ or $C_n^2$ increases, the strength of the turbulence increases. Specifically, the turbulence attenuation at $1\,\textrm{km}$ and $C_n^2=10^{-13}\,\textrm{m}^{-2/3}$ is about $1\,\textrm{dB}$. 
In the THz band with frequency less than $1\,\textrm{THz}$, the attenuation caused by turbulence within $10\,\textrm{km}$ is less than $10\,\textrm{dB}$.
	\section{Conclusion}\label{sec:concl}
In this paper, we have investigated the attenuation and LoSC effect for THz UAV UM-MIMO system in turbulent media.
Specifically, the THz UM-MIMO system model in turbulent media is first established, where the statistical model of atmospheric turbulence is described and the RISC in the THz band is analyzed. 
Second, the effect of LoSC led by turbulent media for UM-MIMO is characterized by calculating the ergodic capacity, which is shown to depend on the NC between different SISO channel gains in the UM-MIMO channel. According to Kolmogorov spectrum of refractive index in turbulent media and Rytov's approximation, the closed-form statistical model for NC and the additional loss of LoSC is derived. 
Third, the fading and attenuation characteristics in the THz band caused by atmospheric turbulence are studied. 
The PDF of the turbulence fading is modeled as a Gamma-Gamma distribution, and the attenuation model based on the Gamma-Gamma fading is derived in a closed-form expression. 
Finally, numerical results demonstrate that When the distance is $10\,\textrm{km}$ and the RISC is $10^{-9}\,\textrm{m}^{-2/3}$, the LoSC effect leads to $10\,\textrm{dB}$ additional loss for a $1024\times1024$ UM-MIMO. Moreover, the turbulence attenuation within $10\,\textrm{km}$ is less than $10\,\textrm{dB}$ in the THz band. These results indicate that the influence of atmospheric turbulence is only important for long-distance ($L\sim 10\,\textrm{km}$) UAV links in strong-turbulence ($C_n^2\sim 10^{-9}\,\textrm{m}^{-2/3}$) environments. 

\begin{appendices}
	\section{Proof of Theorem 1}\label{ap1}
By adopting~\eqref{eq:Uiq/Ui0}, the term $E_{i,0}(\textbf{s},z)/E_{i,0}(\textbf{r}_j,L)$ can be expressed as
\begin{equation}
\begin{aligned}
	&\frac{E_{i,0}(\textbf{s},z)}{E_{i,0}(\textbf{r}_j,L)}=\frac{L}{z}\exp\Big[-\textrm{i}k(L-z)+\textrm{i}k\Big(\frac{s^2}{2z}-\frac{r_j^2}{2L}\Big)\\
    &+\frac{\textrm{i}kr_i^2}{2}\left(\frac{1}{z}-\frac{1}{L}\right)-\textrm{i}k\left(\frac{\textbf{s}\cdot\mathbf{r}_i}{z}-\frac{\mathbf{r}_i\cdot\mathbf{r}_j}{L}\right)\Big],
\end{aligned}
	\label{eq:Ui0/Ui0}
\end{equation}
where $s=|\mathbf{s}|$, $r_i=|\mathbf{r}_i|$, and $r_j=|\mathbf{r}_j|$.
Then, by substituting~\eqref{eq:Ui0/Ui0} and~\eqref{eq:n1_integ} into~\eqref{eq:Uiq/Ui0_L}, the first-order linear perturbation of electric field is given by
\begin{align}
\notag&\widetilde{E}_{i,1}(\mathbf{r}_j,L)\\
\notag=&~\frac{k^2}{2\pi}\int_{0}^{L}\d z\iint_{\infty}^{\infty}\frac{\d \upsilon(\mathbf{k},z)}{L-z}\iint_{-\infty}^{\infty}\exp(i\mathbf{k}\cdot\textbf{s})\d s^2\exp\Big[\\
\notag&\times\textrm{i}k(L-z)+\frac{\textrm{i}k|\textbf{r}_j-\textbf{s}|^2}{2(L-z)}\Big]\frac{L}{z}\cdot\exp\Big[-\textrm{i}k(L-z)\\
\notag&+\textrm{i}k\Big(\frac{s^2}{2z}-\frac{r_j^2}{2L}\Big)+\frac{\textrm{i}kr_i^2}{2}\left(\frac{1}{z}-\frac{1}{L}\right)-\textrm{i}k\Big(\frac{\textbf{s}\cdot\mathbf{r}_i}{z}\\
\notag&-\frac{\mathbf{r}_i\cdot\mathbf{r}_j}{L}\Big)\Big]\\
\notag=&~\frac{k^2}{2\pi}\int_{0}^{L}\d z\iint_{\infty}^{\infty}\frac{L}{z}\frac{\d \upsilon(\mathbf{k},z)}{L-z}\exp\Big[\frac{\textrm{i}kr^2}{2(L-z)}+\frac{-\textrm{i}kr^2}{2L}\\
\notag&+\frac{\textrm{i}kr_i^2}{2z}-\frac{\textrm{i}kr_i^2}{2L}+\frac{\textrm{i}k\mathbf{r}_i\cdot\mathbf{r}_j}{2L}\Big]\iint_{-\infty}^{\infty}\exp\Big[\textrm{i}\Big(\mathbf{k}+\frac{\textrm{i}k\mathbf{r}_i}{L-z}\\
&-\frac{\textrm{i}k\mathbf{r}_j}{z}\Big)\cdot\textbf{s}\Big]\exp\left[\frac{\textrm{i}ks^2}{2(L-z)}+\frac{-\textrm{i}ks^2}{2z}\right]\d s^2. 
\label{eq:Ui1/Ui0} 
\end{align}
Then we apply the formulas
\begin{equation}
	\iint_{-\infty}^{\infty}\exp\left[\textrm{i}\mathbf{v}\cdot\mathbf{s}\right]\exp(\textrm{i}us^2)\d s^2=\frac{\pi \textrm{i}}{u}\exp\left[\frac{-\textrm{i}}{4u}\mathbf{v}^2\right],
\end{equation}
where $\mathbf{v}$ and $u$ represent a real vector and real number, respectively. As shown in~\eqref{eq:Ui1/Ui0}, by substituting $u=\frac{\textrm{i}ks^2}{2(L-z)}+\frac{-\textrm{i}ks^2}{2z}$ and $\mathbf{v}=\mathbf{k}+\frac{\textrm{i}k\mathbf{r}_i}{L-z}-\frac{\textrm{i}k\mathbf{r}_j}{z}$,~\eqref{eq:Ui1/Ui0} can be further simplified as
\begin{equation}
\begin{aligned}
	&\widetilde{E}_{i,1}(\mathbf{r}_j,L)=\textrm{i}k\int_{0}^{L}\d z\iint_{-\infty}^{\infty}\d \upsilon(\mathbf{k},z)\\
	&\cdot\exp\Big[-\frac{\textrm{i}z(L-z)}{2kL}\kappa^2-\frac{\textrm{i}z}{L}\mathbf{k}\cdot\textbf{r}_j+
	\frac{\textrm{i}(L-z)}{L}\mathbf{k}\cdot\mathbf{r}_i\Big].
\end{aligned}
\end{equation}
which proves the result.$\hfill\blacksquare$
	\section{Proof of Theorem 2}\label{ap2}
By replacing the first-order linear perturbation of electric field~\eqref{eq:Ui1/Ui0} into~\eqref{eq:Uiq/Ui0}, $\widetilde{E}_{i,2}(\mathbf{r}_j,L)$ can be expressed as
\begin{align}
\notag&\widetilde{E}_{i,2}(\mathbf{r}_j,L)\\
\notag=&~\frac{k^2}{2\pi}\int_{0}^{L}\d z\iint_{-\infty}^{\infty}\d^2s\exp\Big[\textrm{i}k(L-z)+\frac{\textrm{i}k|\mathbf{s}-\mathbf{r}_j|^2}{2(L-z)}\Big]\\
\notag&\times\frac{E_{i,0}(\mathbf{s},z)}{E_{i,0}(\mathbf{r}_j,L)}\widetilde{E}_{i,1}(\mathbf{s},z)\frac{n_1(\mathbf{s},z)}{(L-z)}\\
\notag=&~\frac{ik^3}{2\pi}\int_{0}^{L}\d z\int_{0}^{z}\d z'\iint_{-\infty}^{\infty}\iint_{-\infty}^{\infty}\frac{\d\nu(\mathbf{k},z)\d\nu(\mathbf{k}',z')}{L-z}\\
\notag&\times\frac{L}{z(L-z)}
\exp\Big[\frac{-\textrm{i}k\mathbf{r}_j^2}{2L}+
\frac{\textrm{i}k\mathbf{r}_i^2}{2}\Big(\frac{1}{z}-\frac{1}{L}\Big)+\frac{\textrm{i}k\mathbf{r}_i\cdot\mathbf{r}_j}{L}\\
\notag&+\frac{\textrm{i}k\mathbf{r}_j^2}{2(L-z)}
-\frac{\textrm{i}\kappa'^2z'(z-z')}{2kz}+\frac{\textrm{i}z'\mathbf{k}'r_i}{z}
\Big]\iint_{-\infty}^{\infty}\d s^2\\
\notag&\times\exp\Big[\textrm{i}\Big(\mathbf{k}+\frac{z-z'}{z}\mathbf{k}'+\frac{\textrm{i}k\mathbf{r}_i}{L-z}-\frac{\textrm{i}k\mathbf{r}_j}{z}\Big)\cdot\textbf{s}\Big]\\
\notag&\times\exp\left[\textrm{i}\Big(\frac{k}{2(L-z)}-\frac{k}{2z}\Big)s^2\right]\\
\notag=&-k^2\int_{0}^{L}\d z\int_{0}^{z}\d z'\iint_{-\infty}^{\infty}\iint_{-\infty}^{\infty}\frac{\d\nu(\mathbf{k},z)\d\nu(\mathbf{k}',z')}{L-z}\\
\notag&\times\exp\Bigg\{
\textrm{i}\Big(\mathbf{k}\frac{L-z}{L}+\mathbf{k}'\frac{L-z'}{L}\Big)\mathbf{r}_i
+ \textrm{i}\Big(\mathbf{k}\frac{z}{L}+\mathbf{k}'\frac{z'}{L}\Big)\mathbf{r}_j\\
\notag&
-\frac{\textrm{i}\kappa^2z(L-z)}{2kL}
-\frac{\textrm{i}\mathbf{k}\cdot\mathbf{k}'z(L-z')}{kL}-\frac{\textrm{i}\kappa'^{2}}{2k}\Big[1-\frac{z'}{z}\\
&-\frac{z'^2}{z}-\frac{z'^2}{L}\Big]\Bigg\},
\label{eq:psi2-1}
\end{align}
where $\kappa'=|\mathbf{k}|$.
We assume that the refractive index in the turbulent media is statistically isotropic and homogeneous. 
Then, we can represent the mean of the inner product of $\d \upsilon(\mathbf{k},z)$ and $\d \upsilon(\mathbf{k}',z')$ in~\eqref{eq:psi2-1} by
\begin{equation}
	\langle \d \upsilon(\mathbf{k},z)\d \upsilon(\mathbf{k}',z')\rangle=F_n(\mathbf{k},|z-z'|)\delta(\mathbf{k}+\mathbf{k}')\d^2\kappa \d^2\kappa',
 \label{eq:dvv+}
\end{equation}
where $F_n(\mathbf{k},z)$ denotes the two-dimensional spatial power spectrum derived from the three-dimensional one as
\begin{equation}
	F_n(\kappa_x,\kappa_y,|z|)\triangleq\int_{-\infty}^{\infty}\Phi_n(\kappa_x,\kappa_y,\kappa_z)\cos(z\kappa_z)\d \kappa_z,
\end{equation}
whose integral is expressed as
\begin{equation}
	\int_{\infty}^{\infty}F_n(\mathbf{k},z)\d z=2\pi\Phi_n(\mathbf{k}).
 \label{eq:Fn}
\end{equation}
By adopting the isotropic of the refractive index fluctuation, we can express $\d^2\kappa=\kappa\d\kappa\d\theta$ and $\mathbf{k}\cdot \mathbf{r}=\kappa r \cos\theta$.
Then by replacing $\eta=\frac{1}{2}(z+z')$ and $\mu=z-z'$ and assuming $z\approx z'\approx \eta$ according to $F_n(\mathbf{k},z)\approx 0$ when $z\gg 0$, $M_{ij,1}$ can be simplified as
\begin{equation}
	M_{ij,1}(L)=-2\pi^2k^2L\int_{0}^{1}\int_{0}^{\infty}\kappa \Phi_n(\kappa,L-L\xi)\d \kappa\d\xi,
\end{equation}
which completes the proof.$\hfill\blacksquare$
	\section{Proof of Theorem 3}\label{ap3}
By substituting~\eqref{eq:psi1} into~\eqref{eq:M2}, the second-order moment of the second kind is expressed as
\begin{equation}
	\begin{aligned}
		&M_{ii'jj',2}(L)=k^2\int_{0}^{L}\d z\int_{0}^{L}\d z'\iint_{-\infty}^{\infty}\iint_{-\infty}^{\infty}\langle \d \upsilon(\mathbf{k},z)\\
		&\d \upsilon^*(\mathbf{k}',z')\rangle\exp\Big[-\frac{\textrm{i}z(L-z)}{2kL}\kappa^2+\frac{\textrm{i}z'(L-z')}{2kL}\kappa'^2-\frac{\textrm{i}z}{L}\\
		&\times\mathbf{k}\cdot\textbf{r}_j+\frac{\textrm{i}z'}{L}\mathbf{k}'\cdot\textbf{r}_{j'}+\frac{\textrm{i}(L-z)}{L}\mathbf{k}\cdot\mathbf{r}_i-\frac{\textrm{i}(L-z')}{L}\mathbf{k}'\cdot\mathbf{r}_{i'}\Big].
	\end{aligned}
	\label{eq:E2mn}
\end{equation}
By assuming the isotropic and homogeneity of the refractive index, we further obtain
\begin{equation}
	\langle \d \upsilon(\mathbf{k},z)\d \upsilon^*(\mathbf{k}',z')\rangle=F_n(\mathbf{k},|z-z'|)\delta(\mathbf{k}-\mathbf{k}')\d^2\kappa \d^2\kappa'
 \label{eq:dvv-}
\end{equation}
Thus, the final result can be expressed by 
\begin{equation}
	\begin{aligned}
		&M_{ii'jj',2}(L)\\
  =&~2\pi k^2\int_{0}^{L}\d \eta\int_{0}^{2\pi}\d\theta\int_{-\infty}^{\infty}\kappa\d\kappa\Phi_n(\mathbf{k})\\
		&\times\exp\Big[-\frac{i\eta\mathbf{k}}{L}(\mathbf{r}_j-\mathbf{r}_{j'})+\frac{\textrm{i}(L-\eta)\mathbf{k}}{L}(\mathbf{r}_i-\mathbf{r}_{i'})\Big]\\
		=&~4\pi k^2L\int_{0}^{1}d\xi\int_{0}^{\infty}\kappa\d\kappa\Phi_n(\mathbf{k})J_0\Big[\kappa\big|(1-\xi)(\mathbf{r}_j-\mathbf{r}_{j'})\\
		&+\xi(\mathbf{r}_i-\mathbf{r}_{i'})\big|\Big],
	\end{aligned}
\end{equation}
where we apply $\int_{0}^{2\pi}\exp(\textrm{i}x\cos\theta)\d\theta=2\pi J_0(x)$ and $\xi=\frac{L-z}{L}$. This result completes the proof.$\hfill\blacksquare$
\section{Proof of Theorem 4}\label{ap4}
The integral in~\eqref{eq:Psi_res} can be calculated by
\begin{align}
	\notag
	\int_{0}^{1}&\int_{0}^{\infty}
	\kappa\Phi_n(\kappa)\bigg\{1-J_0\big[\kappa\xi\Delta\mathbf{r}_{jj'}	+\kappa(1-\xi)\Delta\mathbf{r}_{ii'}\big]\bigg\}
	\d\kappa\d\xi\\
	=\,
	\notag
	&0.033C_n^2\int_{0}^{1}\int_{0}^{\infty}
	\kappa^{-\frac{8}{3}}\bigg\{1-J_0\big[\kappa\xi\Delta\mathbf{r}_{jj'}	+\kappa(1-\xi)\\
	&\Delta\mathbf{r}_{ii'}\big]\bigg\}
	\d\kappa\d\xi
	\label{eq:ap4-1}
	\\
	=\, &
	\notag
	-0.033C_n^2\cdot\frac{\Gamma(-\frac{5}{6})}{\Gamma(\frac{11}{6})}\left(\frac{1}{2}\right)^{8/3}\int_{0}^{1}[\xi\Delta\mathbf{r}_{jj'}-(1-\xi)\\
	&\times\Delta\mathbf{r}_{ii'}]^{\frac{5}{3}}\d\xi
	\label{eq:ap4-2}
	\\
	=\, &-0.033C_n^2\cdot\frac{\Gamma(-\frac{5}{6})}{\Gamma(\frac{11}{6})}\left(\frac{1}{2}\right)^{8/3}\frac{3}{8}\frac{\Delta\mathbf{r}_{ii'}^{\frac{8}{3}}-\Delta\mathbf{r}_{jj'}^{\frac{8}{3}}}{\Delta\mathbf{r}_{ii'}-\Delta\mathbf{r}_{jj'}}
	\label{eq:ap4-3}
\end{align}
where~\eqref{eq:ap4-1} is expanded by replacing $\Phi(\kappa)$ according to~\eqref{eq:Phi_Kolmo}.
Equation~\eqref{eq:ap4-2} is achieving by using the property of Bessel function as
\begin{equation}
\int_{0}^{\infty}\kappa^{-8/3}\Big[1-J_0(b\kappa)\Big]\d\kappa=-\frac{\Gamma(-\frac{5}{6})}{\Gamma(\frac{11}{6})}\left(\frac{1}{2}\right)^{8/3}b^{5/3}.
\end{equation}
Therefore, when $\Delta\mathbf{r}_{ii'}\neq\Delta\mathbf{r}_{jj'}$, we achieve the result as
\begin{equation}
\begin{aligned}
\rho_{ij,ij'}=&~\exp\Bigg(-4\pi^2k^2L \cdot0.033C_n^2\cdot\frac{\Gamma(-\frac{5}{6})}{\Gamma(\frac{11}{6})}\left(\frac{1}{2}\right)^{8/3}\\
&\times\frac{3}{8}\frac{\Delta\mathbf{r}_{ii'}^{\frac{8}{3}}-\Delta\mathbf{r}_{jj'}^{\frac{8}{3}}}{\Delta\mathbf{r}_{ii'}-\Delta\mathbf{r}_{jj'}}\Bigg)\\
\approx&~\exp\left(-0.546C_n^2k^2L\cdot\frac{\Delta\mathbf{r}_{ii'}^{\frac{8}{3}}-\Delta\mathbf{r}_{jj'}^{\frac{8}{3}}}{\Delta\mathbf{r}_{ii'}-\Delta\mathbf{r}_{jj'}}\right).
\end{aligned}
\end{equation}
Otherwise, when $\Delta\mathbf{r}_{ii'}=\Delta\mathbf{r}_{jj'}$, the result becomes
\begin{equation}
\rho_{ij,ij'}\approx\exp\left(-1.457C_n^2k^2L\Delta\mathbf{r}_{ii'}^{5/3}\right),
\end{equation}
which completes the proof.$\hfill\blacksquare$
\end{appendices}
	\bibliographystyle{IEEEtran}
	\bibliography{main}
\end{document}